\documentclass[final,5p,times,twocolumn,authoryear]{elsarticle}

\usepackage{hyperref}
\usepackage{cleveref}

\usepackage{graphicx}
\usepackage{url}
\usepackage{color}

\usepackage{amssymb}

\usepackage{natbib}

\journal{Journal of Atmospheric and Solar-Terrestrial Physics}

\begin{document}

\begin{frontmatter}



\title{Multi-instrument view on solar eruptive events observed with the Siberian Radioheliograph:
From detection of small jets up to development of a shock wave and
CME}


\author{V.V.~Grechnev, S.V.~Lesovoi, A.A.~Kochanov, A.M.~Uralov, A.T.~Altyntsev,
A.V.~Gubin, D.A.~Zhdanov, E.F.~Ivanov, G.Ya.~Smolkov, L.K.~Kashapova}

\address{Institute of Solar-Terrestrial Physics, Irkutsk, Russia}

\begin{abstract}

The first 48-antenna stage of the Siberian Radioheliograph (SRH)
started single-frequency test observations early in 2016, and
since August 2016 it routinely observes the Sun at several
frequencies in the 4--8 GHz range with an angular resolution of
1--2 arc minutes and an imaging interval of about 12 seconds. With
limited opportunities of the incomplete antenna configuration, a
high sensitivity of about 100~Jy allows the SRH to contribute to
the studies of eruptive phenomena along three lines. First, some
eruptions are directly visible in SRH images. Second, some small
eruptions are detectable even without a detailed imaging
information from microwave depressions caused by screening the
background emission by cool erupted plasma. Third, SRH
observations reveal new aspects of some events to be studied with
different instruments. We focus on an eruptive C2.2 flare on 16
March 2016 around 06:40, one of the first flares observed by the
SRH. Proceeding from SRH observations, we analyze this event using
extreme-ultraviolet, hard X-ray, white-light, and metric radio
data. An eruptive prominence expanded, brightened, and twisted,
which indicates a time-extended process of the flux-rope formation
together with the development of a large coronal mass ejection
(CME). The observations rule out a passive role of the prominence
in the CME formation. The abrupt prominence eruption impulsively
excited a blast-wave-like shock, which appeared during the
microwave burst and was manifested in an ``EUV wave'' and Type II
radio burst. The shock wave decayed and did not transform into a
bow shock because of the low speed of the CME. Nevertheless, this
event produced a clear proton enhancement near Earth. Comparison
with our previous studies of several events confirms that the
impulsive-piston shock-excitation scenario is typical of various
events.

\end{abstract}


\end{frontmatter}


\section{Introduction}
\label{S-introduction}

Solar flares, coronal mass ejections (CMEs), associated shock
waves, and related phenomena are known as causes of space weather
disturbances. Hard electromagnetic emissions and energetic
particles pose hazard to space-borne equipment, astronauts on
spacecraft, and even crew members and passengers on aircraft that
carry out transocean flights entering high latitudes.
CME-associated shock waves travel over large distances in the
heliosphere, being responsible for the geomagnetic storm sudden
commencement (SSC). Magnetic structures of CMEs hitting the
Earth's magnetosphere can cause strong geomagnetic storms.

In spite of a certain space weather impact, the origin and
interrelation of solar eruptive phenomena are still not quite
clear. Comprehending solar eruptions is hampered by observational
difficulties. The existing concepts are mostly based on the
hypotheses that were proposed several decades ago and
back-extrapolated results of in-situ measurements in near-Earth
space.

According to a widely accepted view, the main driver of a solar
eruption is a magnetic flux rope. It is considered as the active
structure of a CME that governs its development and subsequent
expansion. The flux rope is traditionally assumed to be associated
with the CME cavity. Prominences (filaments) or associated
structures appear to be among the most probable flux-rope
progenitors \citep{Gibson2015}. However, genesis of flux ropes,
their size range, and other properties are not clear so far.
According to some concepts, the flux rope pre-exists before the
eruption onset \citep{Chen1989, Chen1996, Cheng2013}. Different
concepts relate the flux-rope formation to reconnection processes,
which are also responsible for solar flares
\citep{InhesterBirnHesse1992, LongcopeBeveridge2007, Qiu2007}.

There is no consensus about coronal shock waves. Some authors
advocate flare-ignited blast waves at least in some events
\citep{Magdalenic2010, Magdalenic2012, Nindos2011}. Different
studies demonstrate the CME-related origin of shock waves to be
more probable (e.g. \citealp{Cliver2004}). While basic excitation
mechanisms of shock waves seem to be known (see, e.g.,
\citealp{VrsnakCliver2008}), observational difficulties result in
large uncertainties in their identification.

Solar eruptions and associated phenomena are manifested in
different spectral domains, including microwaves. Radio emission
is produced by various mechanisms, providing important information
on these phenomena and responsible processes. Being sensitive to
gyrosynchrotron emission of nonthermal electrons, microwaves
reveal the flare regions. The microwave spectrum contains
information about accelerated electrons and magnetic fields in the
corona. Being sensitive to thermal plasma emission, microwave
images show eruptive prominences (filaments). Screening background
solar emission by erupted prominence material sometimes produces
depressions detectable even in the total microwave flux
\citep{CovingtonDodson1953} termed the ``negative bursts''. From
studies of the negative bursts, events with reconnection between
erupting structures and a large-scale coronal magnetic environment
were identified \citep{Grechnev2013neg, Grechnev2014_I,
Uralov2014}. These examples demonstrate significant contribution
to studies of solar eruptions from microwave imaging and
non-imaging observations.

Microwave images produced by radio heliographs generally have a
poorer spatial resolution relative to extreme-ultraviolet (EUV)
and X-ray telescopes. Nevertheless, sometimes it is even possible
to judge about the structures that are unresolved in microwave
images \citep{GrechnevKochanov2016, Grechnev2017_II, Lesovoi2017}.

In 2016, the first 48-antenna stage of the Siberian
Radioheliograph (SRH; \citealp{Lesovoi2014, Lesovoi2017}) started
observing the Sun. An overview of the SRH data has revealed
several indications of eruptions. Proceeding from these
indications, we consider a few eruptive events observed by
different instruments and endeavor to address the challenges
listed in this section. We pay special attention to the 16 March
2016 eruptive event, one of the first flares observed by the SRH
\citep{Lesovoi2017}. Multi-instrument analysis of large-scale
aspects of this event promises shedding additional light on the
development of a CME and associated shock wave.

Section~\ref{S-srh} outlines the SRH. Section~\ref{S-neg_bursts}
presents observations of microwave depressions caused by small
jets. Section~\ref{S-eruption_may_1} presents direct observations
of a spray on 1 May 2017. Section~\ref{S-march16} is devoted to a
multi-instrument analysis of an eruptive event on 16 March 2016
that produced a CME and caused a near-Earth proton enhancement.
Section~\ref{S-discussion} discusses the results and shows their
relevance to a typical eruptive event. Section~\ref{S-summary}
summarizes our conclusions and their implications and presents
last changes in the functionality of the SRH.

\section{SRH: 48-Antenna First Stage}
\label{S-srh}

The SRH was constructed as an upgrade of the Siberian Solar Radio
Telescope (SSRT: \citealp{Smolkov1986, Grechnev2003ssrt}). The SSRT
was designed as a cross-shaped interferometer comprising two linear
arrays in the EW and SN directions, each with 128 equidistant
antennas of 2.5\,m diameter spaced by $d = 4.9$\,m. The SSRT scans
the Sun due to its diurnal passage through the fan beam formed by
the simultaneous receiving at a number of different but close
frequencies in the 5.67--5.79\,GHz band. Thus, the SSRT can produce
the images practically at a single frequency every 2--3 minutes at
most.

\begin{figure} 
   \centerline{\includegraphics[width=0.48\textwidth]
    {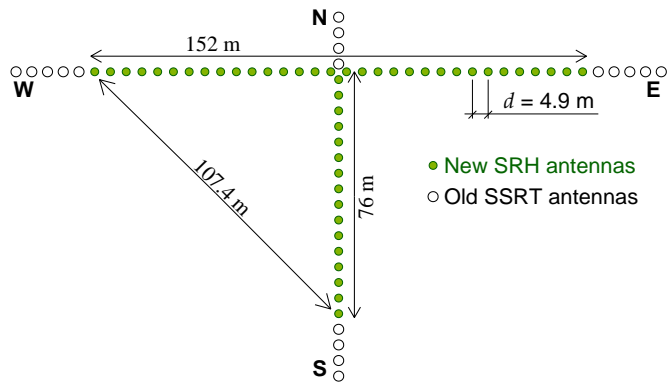}
   }
   \caption{The T-shaped configuration of the 48-antenna SRH first stage.
The remote parts of the four SSRT arms (each arm of 311~m) with
remaining old antennas are not shown. }
   \label{F-SRH_config}
\end{figure}

\begin{figure*} 
   \centerline{\includegraphics[width=\textwidth]
    {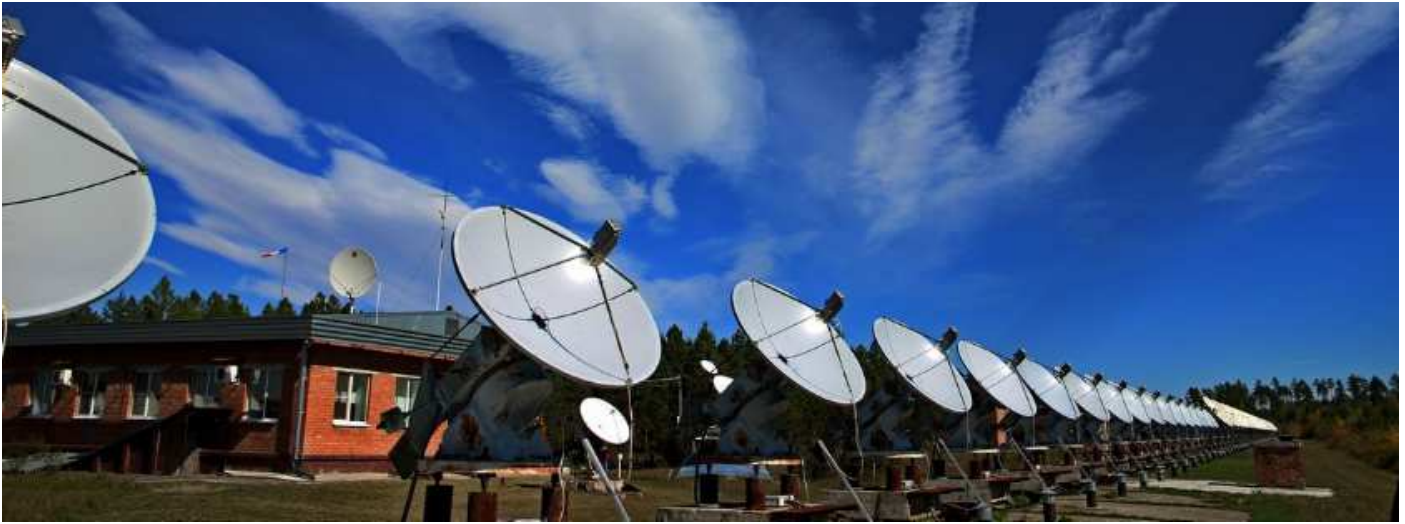}
   }
   \caption{View of the 48-antenna SRH first stage (the east arm).
White remote larger dishes on the right (east) belong to the old
SSRT antenna system. Separate dishes on the ground behind the SRH
antennas belong to the total-flux spectropolarimeters
\citep{ZhdanovZandanov2015}. The receiver and control systems are
located in the working building visible behind the SRH antennas on
the left.}
   \label{F-heliograph}
\end{figure*}

Unlike the directly-imaging SSRT, the SRH uses the Fourier
synthesis. The temporal resolution determined by the receiver
system is much higher than the SSRT had. The SRH has a T-shaped
antenna array. Its 1.8\,m antenna elements replace old SSRT
antennas, being installed at the existing posts along the east,
west, and south arms. The first 48-antenna stage constitutes a
dense equidistant part of a future complete SRH antenna array
(Figures \ref{F-SRH_config} and \ref{F-heliograph}). Being
redundant, this array provides a high sensitivity, which is about
1000~K in the images and reaches for compact sources  $10^{-4}$ of
the total solar flux, i.e. about 100~Jy \citep{LesovoiKobets2017}.

Both circularly-polarized components are measured. The observing
frequencies, each of the 10\,MHz bandwidth in the 4--8\,GHz range,
are set by software and can be optimized for an observing program.
The accumulation time at each frequency is 0.28\,s for each
circularly-polarized component, and the time to switch from one
frequency to another was about 2\,s in 2016 and 2017. The maximum
baseline used is 107.4\,m, enabling a spatial resolution down to
$70^{\prime \prime}$ at 8\,GHz.

\begin{figure} 
   \centerline{\includegraphics[width=0.48\textwidth]
    {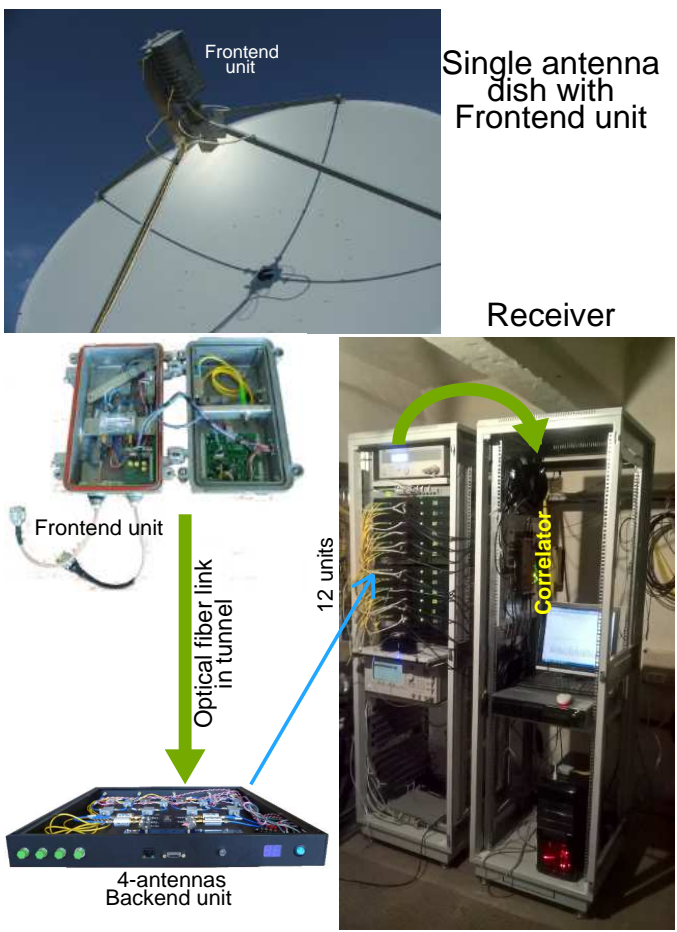}
   }
   \caption{A scheme of the SRH hardware. The frontend units (middle left) are
installed in all antenna elements (top left). The backend of the
receiver and the correlator (bottom right) are located in the
working building. Twelve backend units are mounted in the left
cabinet, and the correlator is located in the right cabinet. The
green arrows denote the paths of the signals. }
   \label{F-SRH_struct}
\end{figure}

The SRH systems outlined in Figure~\ref{F-SRH_struct} were mostly
developed and constructed by the SRH team. The top image represents
a single antenna element. The antenna feed receives two orthogonal
linearly-polarized signals, which come into the frontend unit. A
3-dB $90^{\circ}$ hybrid coupler performs the linear-to-circular
polarization conversion of the input signals. Then they are
pre-amplified and come to a switch, which alternately passes the
left-handedly polarized signal (LCP) and the right-handedly
polarized one (RCP). The signals from the output of the switch come
through the second amplifier to a diode laser, which converts the
ultrahigh-frequency (UHF) signals to optical signals for their
transmission to the working building. The total gain of the frontend
unit is 30--40\,dB.

The signal from each antenna element is transmitted to the backend
of the receiver located in the working building
(Figure~\ref{F-heliograph}) through the optical fiber link located
in the tunnel. Each backend unit (Figure~\ref{F-SRH_struct},
bottom-left) processes the signals from four antennas. The input
optical signals are converted back to the UHF, amplified,
transformed to an intermediate frequency, and digitized at
100\,MHz. Their subsequent digital processing includes the
formation of the operating frequency band, coarse and fine
compensation for the geometric delays and difference in the cable
lengths, and fringe stopping. Finally the digital signals come to
a correlator mounted in the right cabinet shown in
Figure~\ref{F-SRH_struct} (bottom right). The correlator currently
produces 512 complex visibilities for the imaging and several tens
of those for the calibration purposes. Redundant baselines are not
used in the imaging.

Single-frequency test observations started at the SRH early in
2016. Since July 2016 till December 2017, the SRH routinely
observed the Sun at five frequencies. To monitor solar activity
and main SRH systems, the so-called correlation plots are used.
Being a proxy of radio flux, they represent temporal variations in
the sum of cross-correlations of all antenna pairs
\citep{LesovoiKobets2017} and show the changes in both the
brightness and structure of the sources. Real-time correlation
plots and quick-look images produced by the SRH at a set of the
operating frequencies are accessible online at the SRH Web site
\url{http://badary.iszf.irk.ru/}. Adjustment of the SRH systems is
still in progress.

Raw SRH data contain complex visibilities measured at a given set
of frequencies in right and left circularly-polarized components,
information on the array geometry, time stamps, etc. The data are
stored in binary FITS tables. The Python-based library providing
basic programming user interfaces for data handling, phase
calibration, and interferometric imaging routines is under
development.

The phase calibration tasks use the baseline redundancy of the
east, west, and south SRH arms and resolve phase ambiguities in a
sense of an overdetermined optimization problem. To clean raw SRH
images, we tentatively apply an MS-CLEAN algorithm
\citep{Cornwell2008}. Parameters of the algorithm would be
adjusted to meet diverse observational requirements.

The technique to calibrate the images in brightness temperatures
\citep{Kochanov2013, Lesovoi2017} is based on a well-known method
by referring to the most frequent pixel values over the solar disk
and those over the sky. We refer the quiet-Sun brightness
temperature to the measurements by \cite{Zirin1991} and
\cite{Borovik1994}, fitting their frequency dependence with a
fourth-order polynomial in the log--log scale. In particular, we
adopt the values of 21.6, 18.1, 16.0, 14.6, and 13.6 thousand
Kelvin at frequencies of 4.0, 5.0, 6.0, 7.0, and 8.0 GHz,
respectively.

The remaining outer SSRT antennas of the three arms and the whole
north arm continue observing in the original operating mode,
providing the images of compact sources at 5.7\,GHz with a
resolution of down to $21^{\prime \prime}$. Daily quick-look SSRT
images near the local noon are available at the SRH Web site.

\section{Microwave Depressions}
\label{S-neg_bursts}

Temporary depressions of the total microwave flux below the
quasi-stationary level known as negative bursts were discovered by
\cite{CovingtonDodson1953} from observations at 10.7\,cm
(2.8\,GHz). Typically, a negative burst follows an ordinary
flare-related impulsive burst, when the eruption screens a radio
source located in the same or a nearby active region. The cause of
a negative burst is screening by low-temperature absorbing erupted
material of a compact microwave source \citep{Covington1973,
Sawyer1977, Maksimov1991} or/and large areas of the quiet Sun.
Hence, microwave depressions indicate probable eruptions. The
dependence of the absorption depth on both the observing frequency
and properties of absorbing plasma provides a basic possibility to
estimate some parameters of an erupting structure, if a depression
is observed at different frequencies (see, e.g.,
\citealp{Grechnev2008, Grechnev2013neg, KuzmenkoGrechnev2017}).

Because both the opacity of a filament or surge and its contrast
against the solar disk depend on the frequency inversely, the
negative bursts are observed mainly at 1--10\,GHz. Although
eruptions occur often, detection of microwave depressions requires a
high sensitivity and calibration stability of total-flux radiometers
that makes the negative bursts rare phenomena. From 1990 through
2009, their total number recorded by all ground-based stations was
72 with a maximum yearly number being as small as 14 in 1991
\citep{Grechnev2013neg}. Previously negative bursts were observed
almost exclusively in total intensity.

With an operating frequency range within 4--8\,GHz and a high
sensitivity, the SRH observations promise the detection of
eruption-related absorption phenomena. A simplest way to detect a
microwave depression is provided by the correlation plots.
\cite{Lesovoi2017} presented an unprecedented series of three
negative bursts observed in one day on 9 August 2016 by the SRH and
Nobeyama Radio Polarimeters (NoRP: \citealp{Torii1979,
Nakajima1985}) in both intensity and polarization. These negative
bursts were caused by repeating surges, which screened a polarized
sunspot-associated microwave source in active region (AR) 12574
located not far from the limb (N04\,E59).

Here we present examples of microwave depressions revealed from
the SRH data that really point at small eruptions. Some of the
eruptions indicated by the SRH are too weak and small to be easily
detected from observations at different wavelengths. The
possibilities of plasma diagnostics for such eruptions based on
the SRH data are discussed in Section~\ref{S-summary_eruptions}.

\subsection{A Small Eruption on 9 September 2017}

\begin{figure} 
   \centerline{\includegraphics[width=0.48\textwidth]
    {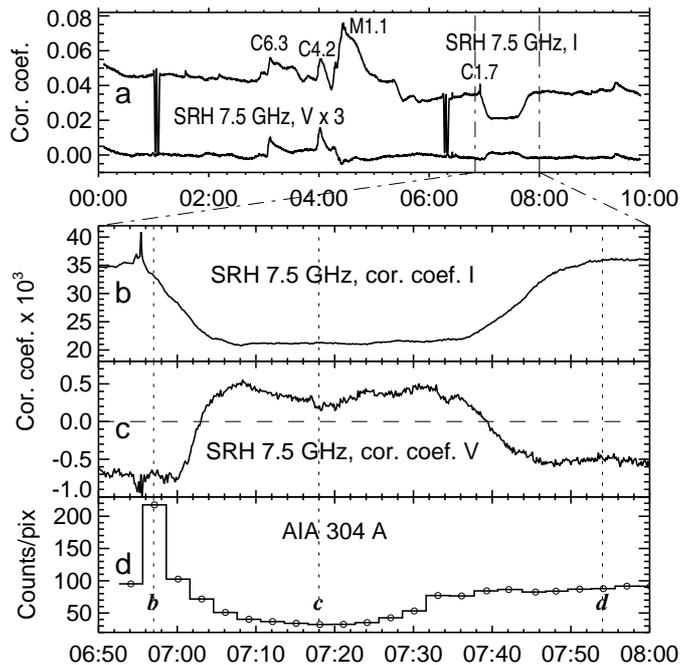}
   }
   \caption{Temporal profiles of the small eruption on 9 September 2017.
The temporal profile in panel d was computed over the framed
region in Figure~\ref{F-2017-09-09_aia304} from the
quarter-resolution beacon AIA 304\,\AA\ images with a 3-minute
interval. The vertical dotted lines denote the times of the images
in Figure~\ref{F-2017-09-09_aia304} whose panels are indicated by
the bold-italic letters in panel d.}
   \label{F-2017-09-09_timeprof}
\end{figure}

A conspicuous microwave depression recorded on 9 September 2017 is
visible between the vertical dash-dotted lines in
Figure~\ref{F-2017-09-09_timeprof}a, which presents the SRH
intensity and polarization correlation plots at a frequency of
7.5\,GHz. The bursts at 03:06, 04:00, 04:26, and a spiky burst at
06:55 are associated with GOES C6.3, C4.2, M1.1, and C1.7 flares,
respectively, all of which occurred in AR\,12673. The excursions
around 01:00 and 06:15 are caused by the Sun-to-sky calibration
maneuvers of the antenna system. The depression in intensity has a
counterpart in polarization, indicating the screening of a
polarized source. The plots at the other frequencies are similar.
The SRH images reveal that the brightness decreased in a microwave
source located close to the west limb.

The depression was caused by a small eruption associated with a
short (7 minutes) impulsive C1.7/1F flare (S10\,W70) in AR\,12673.
This superactive region produced from 4 through 10 September four
X-class flares and numerous weaker events. The major eruptive
events in this region caused strong fluxes of energetic particles,
a severe geomagnetic storm on 7--9 September, a deep Forbush
decrease, and a ground-level enhancement of cosmic-ray intensity
(GLE73) on 10 September, as AR\,12673 arrived at the west limb.
The event of interest was much weaker.

\begin{figure} 
   \centerline{\includegraphics[height=0.9\textheight]
    {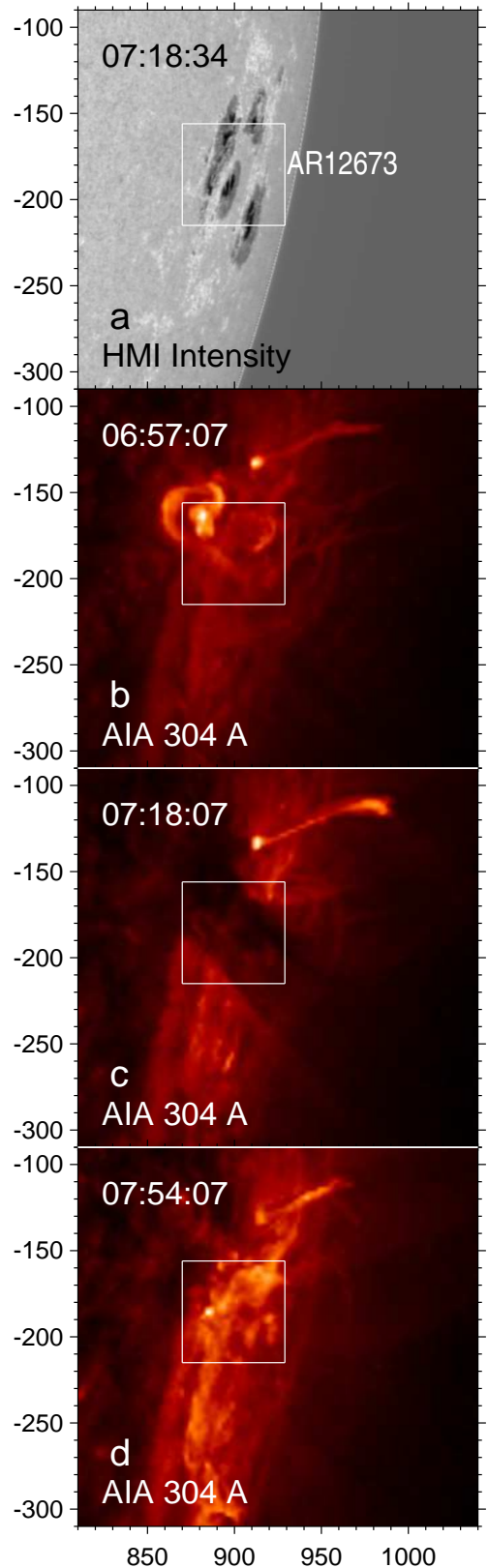}
   }
   \caption{Small eruption on 9 September 2017 in the
SDO/AIA 304\,\AA\ images (b--d) in comparison with a sunspot group
visible in an HMI intensitygram (a). The axes indicate the
distance from solar disk center in arcseconds.}
   \label{F-2017-09-09_aia304}
\end{figure}

The intensitygram in Figure~\ref{F-2017-09-09_aia304}a produced on
9 September by the \textit{Helioseismic and Magnetic Imager} (HMI:
\citealp{Scherrer2012}) onboard the \textit{Solar Dynamics
Observatory} (SDO) shows that AR\,12673 comprised several
sunspots. It had a complex magnetic $\beta \gamma
\delta$-configuration. Figures
\ref{F-2017-09-09_aia304}b\,--\,\ref{F-2017-09-09_aia304}d present
three episodes of the small event observed by the
\textit{Atmospheric Imaging Assembly} (AIA: \citealp{Lemen2012})
onboard SDO in the 304\,\AA\ channel, which is most sensitive to
low-temperature plasma. Here we used quarter-resolution beacon AIA
files available with an interval of 3 minutes. AIA did not observe
the whole Sun between 06:27 and 06:54.
Figure~\ref{F-2017-09-09_aia304}b shows a flare brightening with a
circular ribbon. Figure~\ref{F-2017-09-09_aia304}c reveals a
jet-like eruption. Figure~\ref{F-2017-09-09_aia304}d presents the
active region after the event.

Figures \ref{F-2017-09-09_timeprof}b and
\ref{F-2017-09-09_timeprof}c show expanded correlation plots in
intensity and polarization. While the structure of the active
region is unresolved by the SRH, the change in the polarization
indicates the screening of one or more sunspot-associated sources
in AR\,12673. Figure~\ref{F-2017-09-09_timeprof}d presents the
average brightness in 304\,\AA\ over the framed region in
Figure~\ref{F-2017-09-09_aia304} to compare the EUV and microwave
observations. The microwave depression lasted somewhat longer than
the jet was visible in the 304\,\AA\ images.

The \url{2017-09-09_AIA304_WL_SRH.mpg} movie presents the course
of the event as observed by AIA in 304\,\AA\ (left) in comparison
with HMI intensitygrams (right). The bottom plot shows the same
304\,\AA\ light curve in Figure~\ref{F-2017-09-09_timeprof}d in
white and the 7.5\,GHz correlation plots in yellow scaled to match
the plotted range. The red vertical line on the plots marks the
current observation time. A short-lived flare brightening visible
in one image is followed by an eruption (surge) from the same
region. The rising material of the surge is initially narrow and
bright that indicates its temperature of order $5 \times
10^{4}$~K. Then the surge broadens, darkens, and screens the
structures behind it. The absorption indicates a temperature of
the erupted material of $ < 10^{4}$~K. The surge partly covers a
sunspot group in AR\,12673 behind it. The screening of microwave
sources above the sunspots causes the depression in total
intensity and change in polarization. After 07:20 the opacity of
the surge gradually decreases that corresponds to the recovery of
the 304\,\AA\ emission flux. The microwave emission recovers
later.

The depression was preceded by a small microwave burst around
06:55 corresponding to the flare brightening. Simultaneously, a
group of metric Type~III bursts was observed from 06:53 to 06:56
extending down to the kilometric range that indicates escape of
accelerated electrons into the interplanetary space. No CME
followed this event.

\subsection{A Microeruption on 3 August 2017}

\begin{figure} 
   \centerline{\includegraphics[width=0.48\textwidth]
    {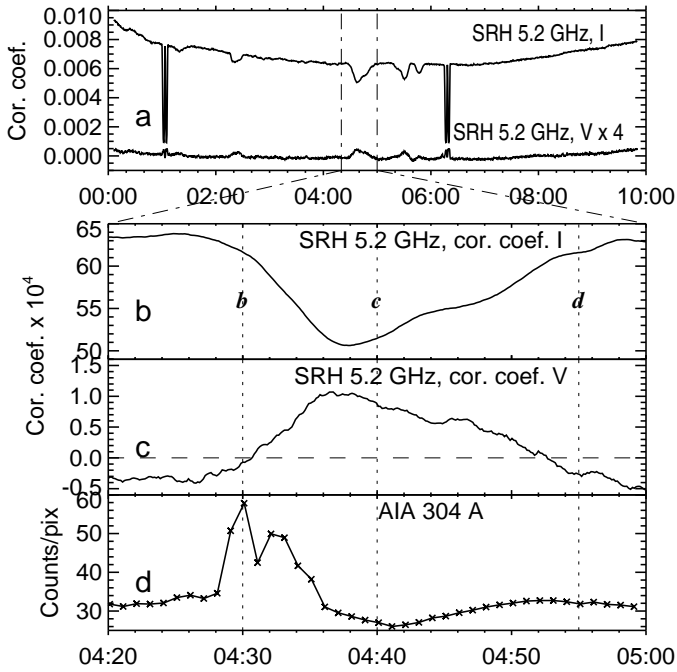}
   }
   \caption{Time profiles of the microeruption on 3 August 2017. The time
profile in panel d was computed over the framed region in
Figure~\ref{F-2017-08-03_aia304} from full-resolution AIA
304\,\AA\ images taken with a 1-minute interval. The vertical
dotted lines denote the times of the images in
Figure~\ref{F-2017-08-03_aia304} whose panels are indicated by the
bold-italic letters in panel b.}
   \label{F-2017-08-03_timeprof}
\end{figure}

\begin{figure} 
   \centerline{\includegraphics[height=0.9\textheight]
    {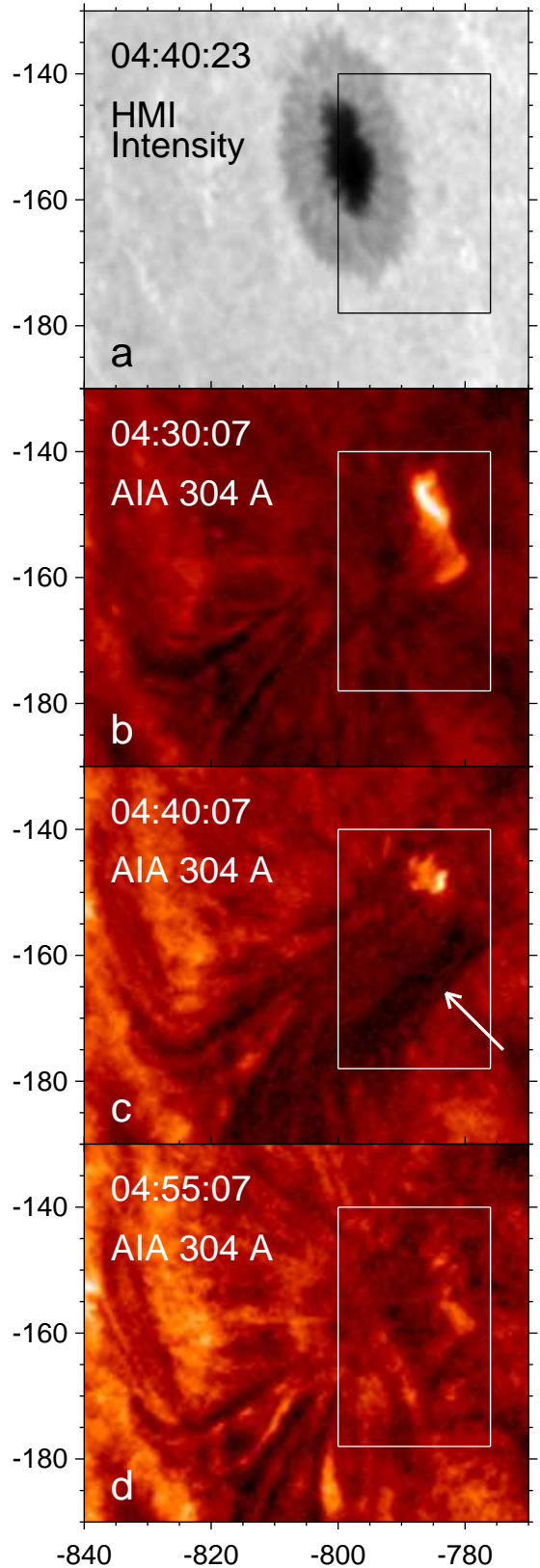}
   }
   \caption{Microeruption on 3 August 2017 in the
SDO/AIA 304\,\AA\ images (b--d) in comparison with a sunspot
visible in an HMI intensitygram (a). The arrow in panel c
indicates a tiny surge. The axes indicate the distance from solar
disk center in arcseconds.}
   \label{F-2017-08-03_aia304}
\end{figure}

A microwave depression caused by a still weaker eruptive event was
observed on 3 August 2017. Figures \ref{F-2017-08-03_timeprof} and
\ref{F-2017-08-03_aia304} present the event occurring in AR\,12670
(S06\,E55) in the formats similar to those in the preceding
section. Note that the SRH correlation coefficients here were one
order of magnitude smaller than in the 9 September 2017 event. To
reduce the noise, they were smoothed in Figures
\ref{F-2017-08-03_timeprof}b and \ref{F-2017-08-03_timeprof}c with
a 15-samples-wide boxcar.

The circumstances of the 3 August and 9 September events are mainly
similar. A brightening visible in 304\,\AA\ near a single isolated
sunspot located not far from the east limb was followed by a tiny
surge (the arrow in Figure~\ref{F-2017-08-03_aia304}c) that
overlapped with a sunspot-associated polarized microwave source and
partly screened its emission.

However, the spatial size and energy of this event were considerably
smaller. The field of view in Figure~\ref{F-2017-08-03_aia304}
roughly corresponds to the SRH beam size, while the region of
brightening is poorly visible even in the full-resolution AIA
304\,\AA\ images. There were no Type~III bursts and no CME. No
response to this event is present in X-rays, and its detection in
AIA images is not a simple task. Nevertheless, this microeruption is
clearly visible in the SRH correlation plots, while its location is
easily identified from the SRH images.

The correlation plots in Figure~\ref{F-2017-08-03_timeprof}a
reveal more depressions on that day. At least one of them, around
02:20, was caused by a similar microeruption in the same active
region. Depressions are also detectable in the SRH data on some
different days.

\section{A Spray Observed on 1 May 2017}
\label{S-eruption_may_1}

\begin{figure} 
   \centerline{\includegraphics[width=0.48\textwidth]
    {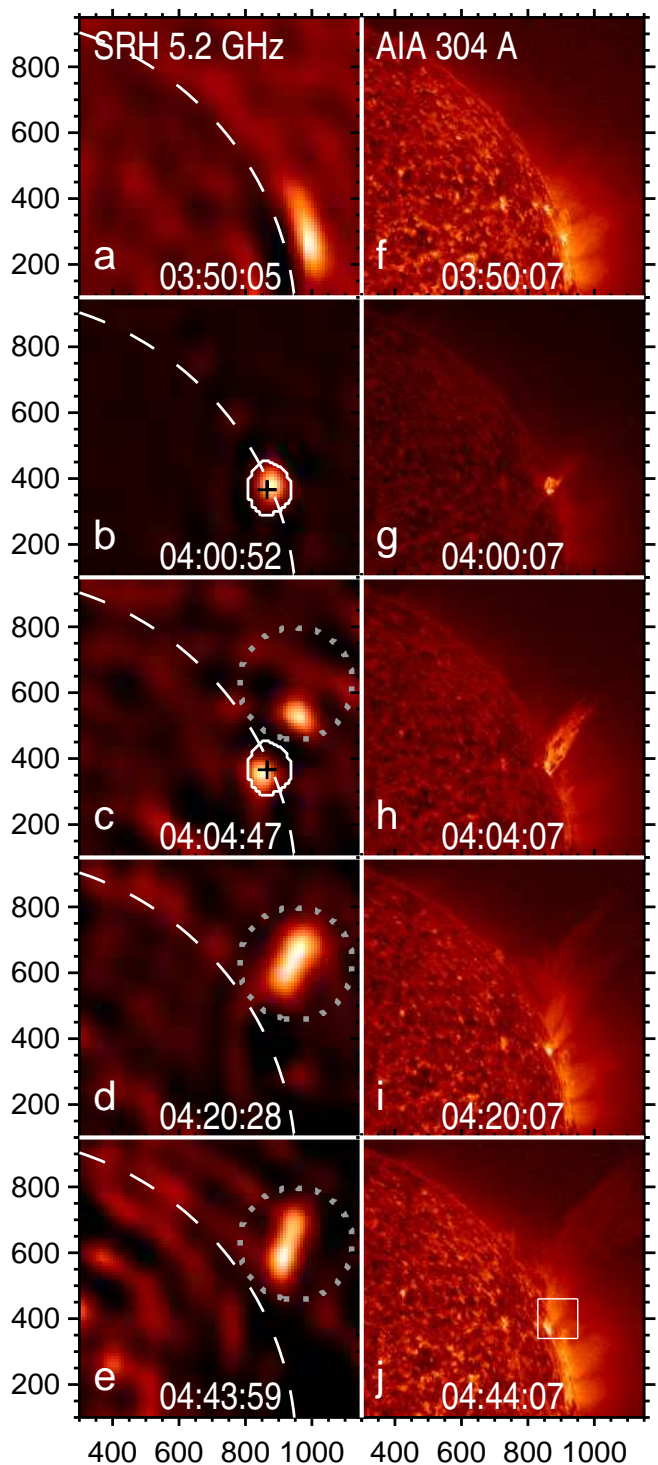}
   }
   \caption{Eruption on 1 May 2017 in dirty SRH 5.2\,GHz images (a--e)
in comparison with SDO/AIA 304\,\AA\ images (f--j). The solar disk
is subtracted in the SRH images and reduced in the AIA images. The
dashed arc denotes the limb. The solid contours in panels {b, c}
outline the flaring source. The gray-dashed circle in panels
{c--e} outline the off-limb eruption. The temporal profiles over
the contoured region are presented in
Figure~\ref{F-2017-05-01_timeprof}a. The black cross in panels {b,
c} denote the center of the X-ray source in RHESSI images. The
frame in panel {j} denotes the field of view in
Figure~\ref{F-2017-05-01_aia211_304}. The axes indicate the
distance from solar disk center in arcseconds.}
   \label{F-2017-05-01_srh_aia304}
\end{figure}

The eruptive event on 1 May 2017 associated with a B9.9 flare in
active region 12652 (N18\,W78) was directly observed by the SRH.
Figure~\ref{F-2017-05-01_srh_aia304} presents the images of the
event produced by the SRH at 5.2\,GHz in the left column along
with temporally close SDO/AIA 304\,\AA\ images in the right
column. Note that the solar disk is subtracted in the SRH images
(the quiet-Sun brightness temperature at 5.2\,GHz is 17570\,K) and
reduced in the 304\,\AA\ images of this event to emphasize the
off-limb spray. The flare region is denoted by the solid contour
in Figures \ref{F-2017-05-01_srh_aia304}b and
\ref{F-2017-05-01_srh_aia304}c. The eruption is outlined in
Figures
\ref{F-2017-05-01_srh_aia304}c--\ref{F-2017-05-01_srh_aia304}e by
the thick gray-dashed circle.

Figures \ref{F-2017-05-01_srh_aia304}a and
\ref{F-2017-05-01_srh_aia304}f show the situation before the
event. A compact flare brightening appears in Figures
\ref{F-2017-05-01_srh_aia304}b and \ref{F-2017-05-01_srh_aia304}g.
A spray appears in the next row
(Figures~\ref{F-2017-05-01_srh_aia304}c,\,h); the SRH shows its
thickest part with a considerably poorer resolution relative to
SDO/AIA. Then, the flaring source disappears at 5.2\,GHz, while a
portion of the off-limb spray is still present in the SRH images.
The spray broadens in 304\,\AA; a part of its material returns to
the solar surface.

The black cross in Figures~\ref{F-2017-05-01_srh_aia304}b,\,c
denotes the brightness center of an X-ray source observed by the
\textit{Reuven Ramaty High-Energy Solar Spectroscopic Imager}
(RHESSI: \citealp{Lin2002}). The centers of the source observed at
3--6\,keV, 6--12\,keV, and 12--25\,keV coincide to within
$2.5^{\prime \prime}$. A response is detectable in the RHESSI
count rate up to the 25--50\,keV band.

\begin{figure} 
   \centerline{\includegraphics[width=0.48\textwidth]
    {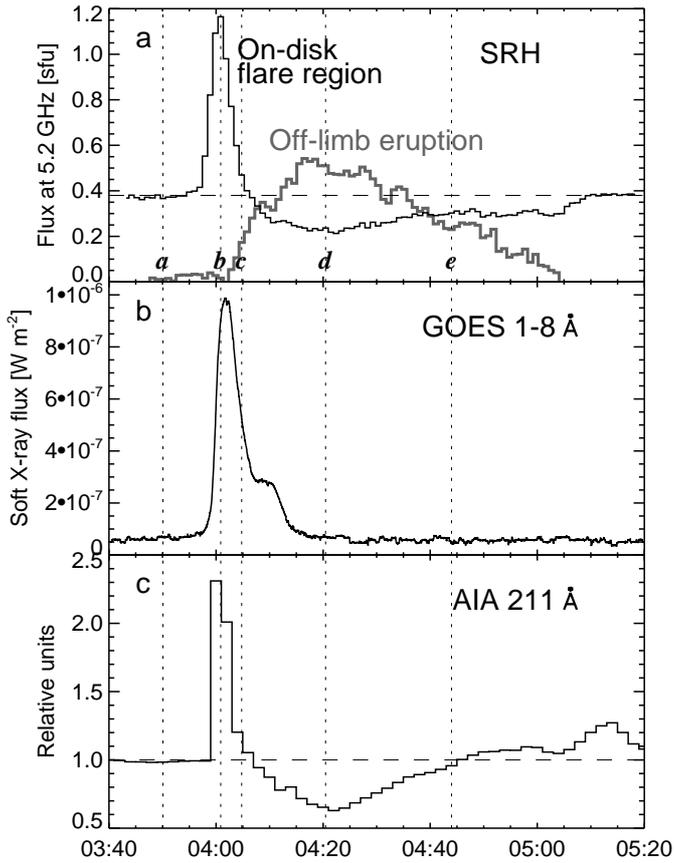}
   }
   \caption{Temporal profiles of the 1 May 2017
eruptive event. a)~Microwave flux profiles computed from the SRH
images at 5.2\,GHz over the flare region (black; solid contour in
Figures~\ref{F-2017-05-01_srh_aia304}b,\,c) and over the off-limb
spray (gray; dotted contour in
Figures~\ref{F-2017-05-01_srh_aia304}c--e). The labels at the
bottom denote the observation times of the corresponding panels in
Figure~\ref{F-2017-05-01_srh_aia304}. b)~GOES 1--8\,\AA\ plot.
c)~Temporal profile computed from the 211\,\AA\ images over a
framed region in Figure~\ref{F-2017-05-01_aia211_304}d.}
   \label{F-2017-05-01_timeprof}
\end{figure}

Figure~\ref{F-2017-05-01_timeprof}a presents the temporal profiles
computed from the SRH images (synthesized with a one-minute
interval) over the contoured regions in comparison with a GOES
1--8\,\AA\ flux shown in Figure~\ref{F-2017-05-01_timeprof}b in
the linear scale. The similarity of the microwave burst (black)
with a soft X-ray (SXR) flux suggests domination of the microwaves
by thermal emission, consistent with a flatness within $\pm 8\%$
of the flux spectrum measured from the SRH images at
4.0--6.8\,GHz. The thermal bremsstrahlung estimated from GOES data
provides 0.8\,sfu equal to the microwave flux actually observed.
The same flux of the microwave source was computed from the
17\,GHz image produced by the Nobeyama Radioheliograph (NoRH:
\citealp{Nakajima1994}) at 04:00. This weak microwave burst is not
detectable in NoRP or Learmonth data. It is only shown by the RT-2
radio telescope of the Ussuriysk Astrophysical Observatory
\citep{Kuzmenko2008} at 2.8\,GHz, where its flux was also about
0.8\,sfu. A flat microwave spectrum over a six-fold frequency
range confirms that the burst was due to optically thin free-free
emission.

\begin{figure} 
   \centerline{\includegraphics[width=0.48\textwidth]
    {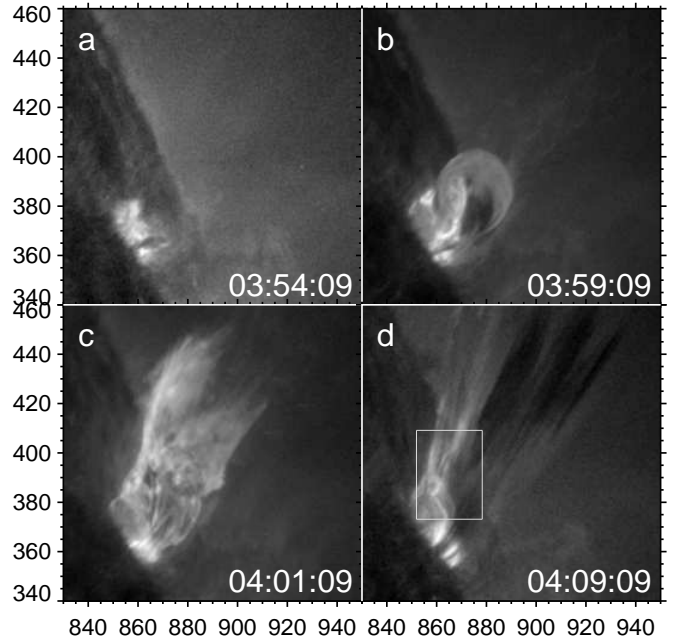}
   }
   \caption{Eruption region on 1 May 2017 in combined
SDO/AIA 304\,\AA\ and 211\,\AA\ images. The times averaged between
both images separated by 4.5\,s are specified in the panels. The
axes indicate the distance from solar disk center in arcseconds.}
   \label{F-2017-05-01_aia211_304}
\end{figure}

Unlike the SXR burst followed by a shoulder, the microwave burst
changed to a depression, which lasted one hour
(Figure~\ref{F-2017-05-01_timeprof}). The depression was most
likely caused by absorption of the microwave emission in the
low-temperature plasma of the spray. A dark absorbing material is
really visible in the combined 211\,\AA\ and 304\,\AA\ AIA images
in Figure~\ref{F-2017-05-01_aia211_304}d. The similarity between
the temporal profile in Figure~\ref{F-2017-05-01_timeprof}c
computed from the 211\,\AA\ images over the framed region and the
microwave profile of the flare region confirms the
absorption-related origin of the microwave depression. The total
microwave flux emitted by the off-limb spray is represented by the
thick-gray line in Figure~\ref{F-2017-05-01_timeprof}a. The
temporal profile of the microwave depression resembles an inverted
profile of the spray that also confirms their common cause.

The filament eruption is seen in combined SDO/AIA 304\,\AA\ and
211\,\AA\ images in Figure~\ref{F-2017-05-01_aia211_304}, whose
field of view is denoted by the frame in
Figure~\ref{F-2017-05-01_srh_aia304}j. A part of a dark
pre-eruptive filament in Figure~\ref{F-2017-05-01_aia211_304}a
screens the bright emission above a plage. In
Figure~\ref{F-2017-05-01_aia211_304}b, a thick circular structure
bound with the filament brightens up. The eruption process
strengthens in Figure~\ref{F-2017-05-01_aia211_304}c corresponding
to the peak of the microwave and X-ray bursts. Two Type~III bursts
occurred at that time extending to the kilometric range that
suggests the appearance of accelerated electrons and their escape
into the interplanetary space. The brightest compact source was
located in the southwest part of the configuration.
Figure~\ref{F-2017-05-01_aia211_304}d shows outflow of
low-temperature plasma along the main legs of the erupting
filament. This plasma partly returned back later. The
low-temperature plasma flow screened the bright microwave-emitting
source that caused the depression in the temporal profile in
Figure~\ref{F-2017-05-01_timeprof}a.

A supplementary \url{2017-05-01_AIA304_SRH.mpg} movie presents the
development of the large-scale spray in 304\,\AA\ images (right) and
the SRH observations at 5.2\,GHz (left). The impulsive flare
brightening is maximum at 04:00. A bright erupted material appears
at 04:02. A dark absorbing low-temperature material appears at 04:04
which corresponds to the decay of the spike at 5.2\,GHz in
Figure~\ref{F-2017-05-01_timeprof}a and in 304\,\AA\ in
Figure~\ref{F-2017-05-01_timeprof}c. The rising motion of the dark
material is visible until 04:14, and then its returning motion
starts. The erupted material visible in 304\,\AA\ gradually falls
until the end of the movie (corresponding to the end of the
depression in Figures~\ref{F-2017-05-01_timeprof}a,\,c), while its
amount decreases.

\begin{figure} 
   \centerline{\includegraphics[width=0.48\textwidth]
    {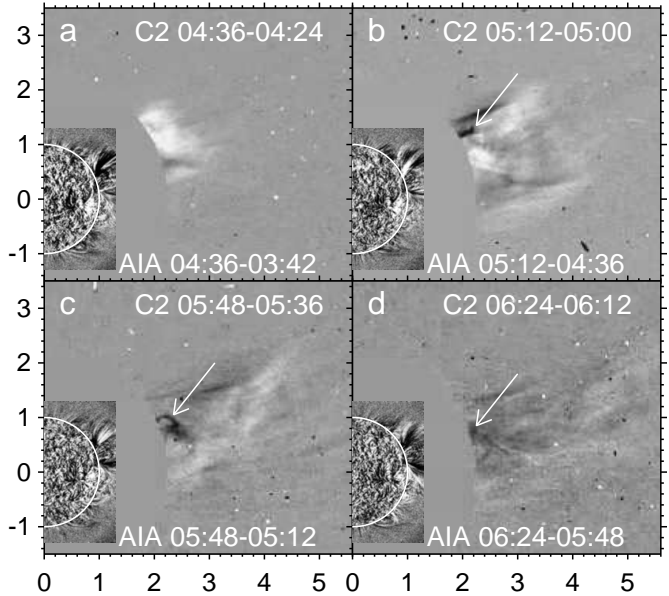}
   }
   \caption{Mass ejection on 1 May 2017 in LASCO-C2 running-difference images.
The insets show the AIA 193\,\AA\ image ratios. The arrows point
at dark features returning to the solar surface. The circles
denote the solar limb. The axes indicate the distance from solar
disk center in solar radii.}
   \label{F-2017-05-01_c2}
\end{figure}

Figure~\ref{F-2017-05-01_c2} shows a mass ejection observed by the
\textit{Large Angle Spectroscopic Coronagraph} (LASCO:
\citealp{Brueckner1995}) onboard SOHO. The ejection also looks
like a spray and does not exhibit a flux-rope-like magnetic
structure. A trailing part of the ejected material (dark in the
running differences) indicated by the arrows returned to the
surface. The ejection dispersed in solar wind and disappeared in
the LASCO-C3 field of view.

\section{The 16 March 2016 Event Associated with a CME and Shock Wave}
 \label{S-march16}

Unlike the small eruptions not associated with CMEs presented in
Section~\ref{S-neg_bursts}, here we consider an eruptive-flare
event, which occurred on 16 March 2016 in AR\,12522 (N14\,W83) and
had a GOES importance of C2.2. The event gave rise to a CME and
shock wave and produced a weak near-Earth proton enhancement. This
was the first flare observed by the SRH, when it operated in a
single-frequency mode at 6.0\,GHz. Here we start from SRH images
and follow different stages of the event using imaging and
non-imaging observations in hard X-rays, extreme ultraviolet,
white-light, and in metric radio range.

\subsection{SRH Observations and Preliminary Conclusions}

\begin{figure} 
   \centerline{\includegraphics[width=0.48\textwidth]
    {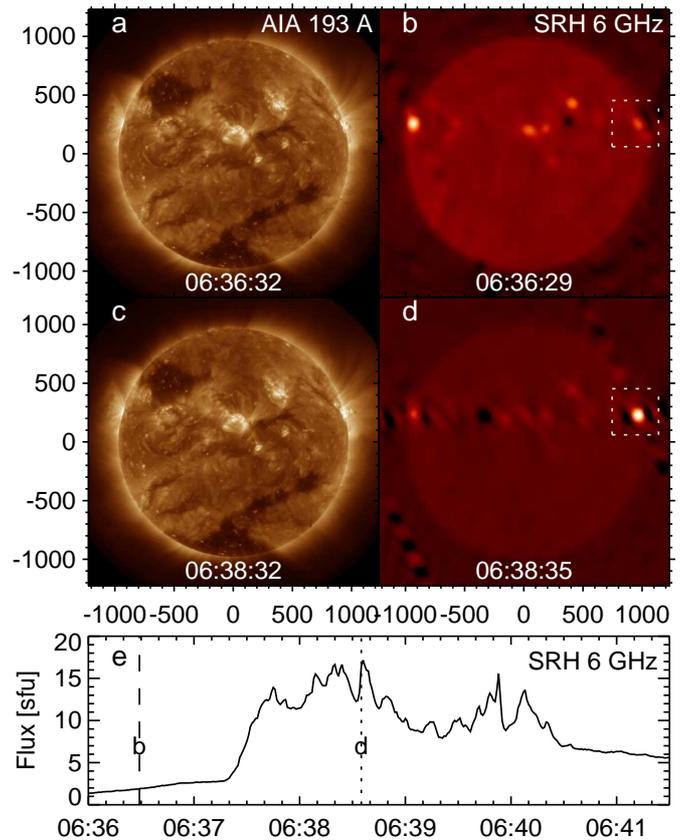}
   }
   \caption{The 16 March 2016 eruptive flare
in the AIA~193\,\AA\ and clean SRH 6\,GHz images: a,\,b)~before
the flare, c,\,d)~near the maximum of the microwave burst, e)~the
total-intensity temporal profile at 6\,GHz computed from the SRH
images over a framed region in panels (b,d).}
   \label{F-2016-03-16_srh_aia}
\end{figure}

We synthesized about 3270 total-intensity (Stokes $I$) images in
steps of 1\,s for the whole flare duration from 06:35:34 to
07:30:10. Each image was processed separately for the impulsive
phase, and we produced 10\,s averages for a later stage. Each of
the images obtained in this way was calibrated in brightness
temperatures individually using the technique described by
\cite{Kochanov2013} and referring to the quiet-Sun brightness
temperature of 15960\,K at 6.0\,GHz. All of the images were
coaligned. One of the images observed by the SRH before the flare
is shown in Figure~\ref{F-2016-03-16_srh_aia}b, and an image
observed close to the maximum of the microwave burst is shown in
Figure~\ref{F-2016-03-16_srh_aia}d. Nearly simultaneous AIA
193\,\AA\ images are shown on the left.

The microwave emission of this flare was too weak to be recorded
by total-flux radiometers properly. With an insufficient spatial
resolution of the SRH to supply detailed images of the flare site,
its sensitivity is high enough to produce a detailed light curve.
The total-flux temporal profile was computed from dirty SRH images
over the flare region denoted by the dotted white frame in
Figures~\ref{F-2016-03-16_srh_aia}b,\,d. The microwave burst was
modest, up to 18\,sfu, while a hard X-ray (HXR) burst was
considerable.

The impulsive phase of the flare is shown by the
\url{2016-03-16_SRH_impulsive_phase_inset.mpg} movie composed from
dirty SRH images with an interval of 1\,s. Each full-disk image is
displayed with an individual nonlinear brightness scale to reveal
the brightness distribution over the solar disk. The top-left
inset represents the framed region in a common linear brightness
scale over the whole flare. The bottom plot shows the total-flux
temporal profile over the framed region with a moving vertical
line, which denotes the observation time of the corresponding
image.

The \url{2016-03-16_AIA193_304_SRH_Fermi.mpg} movie presents the
prominence eruption observed by AIA in 193\,\AA\ (left) and in
304\,\AA\ (right) in comparison with the microwave and HXR bursts
shown at the bottom. The eruption started first; the bursts became
considerable, when intermittent brightenings appeared in 193\,\AA\
near the solar surface beneath the rising prominence. The temporal
structure of the microwave burst is similar to a temporal profile
computed from the running-difference 193\,\AA\ images over
combined regions of the intermittent brightenings, whereas no
similarity was observed with any of the individual regions
\citep{Lesovoi2017}.

SRH images indicate an expanding feature above the west limb. At
that time, the image of the Sun from an adjacent interference
order of the SRH was located close to the main image right on the
west, where the erupting prominence expanded. The east--west
sidelobes from the flare region and those from a source at the
east limb overlapped (Figure~\ref{F-2016-03-16_srh_aia}d),
covering the erupting prominence. Unfavorable observation
conditions and a low contrast of the erupting prominence
determined by a large area of the SRH beam make its analysis from
SRH images difficult. We therefore consider EUV observations of
the erupting prominence in the next section.

A brief analysis of the flare observed by the SRH and the
prominence eruption led \cite{Lesovoi2017} to the following
conclusions: 1.~Acceleration of most electrons in the flare was
initiated by the prominence eruption. 2.~Compact microwave sources
were located in the legs of the flare arcade throughout its whole
length. 3.~HXR sources were most likely also distributed over the
flare ribbons.

Here we continue with a study of this event, focusing on its
large-scale aspects and using data of different instruments. We
also pay attention to its space weather impact.

\subsection{Prominence Eruption}
 \label{S-prom_eruption}

\begin{figure} 
   \centerline{\includegraphics[height=0.9\textheight]
    {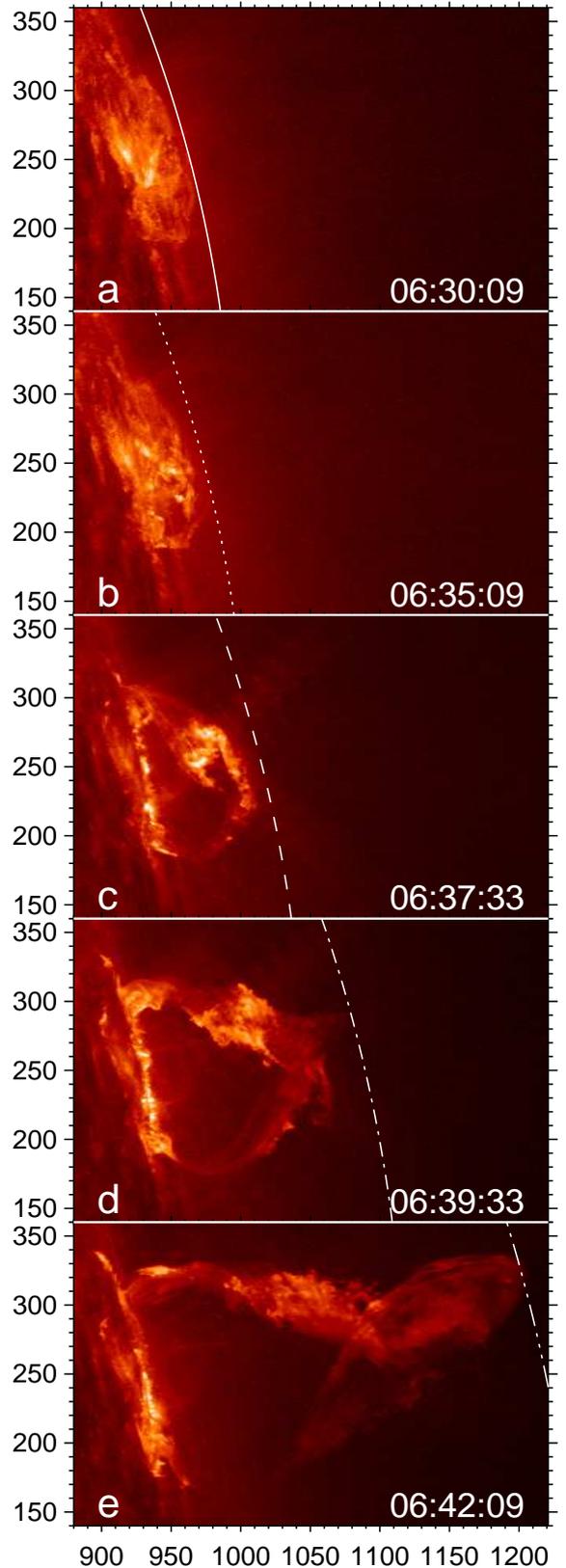}
   }
   \caption{Prominence eruption on 16 March 2016 in the
SDO/AIA 304\,\AA\ images. The arcs outlining the top of the
erupting prominence correspond to the kinematic curves presented
in Figure~\ref{F-kinem}. The axes indicate the distance from solar
disk center in arcseconds.}
   \label{F-2016-03-16_aia304}
\end{figure}

AIA 304\,\AA\ images in Figure~\ref{F-2016-03-16_aia304} present
some episodes of the prominence eruption.
Figure~\ref{F-2016-03-16_aia304}a shows the initial static
prominence. In Figure~\ref{F-2016-03-16_aia304}b, the southern
part of the prominence top slightly displaced up, and a gap in its
body appeared beneath. Flare ribbons are not yet detectable. In
Figure~\ref{F-2016-03-16_aia304}c, the prominence considerably
stretched up. Its broadest part north of the top brightened up,
that indicates heating; note faint cross-shaped diffraction
patterns on the photodetector emanating from this bright feature.
A flare ribbon appeared. In Figure~\ref{F-2016-03-16_aia304}d, the
prominence changed still stronger, having acquired a high speed.
The top part took a complex shape and started stretching forward.
In Figure~\ref{F-2016-03-16_aia304}e, the twisted prominence
intersected. Two ribbons are visible.

\begin{figure} 
   \centerline{\includegraphics[width=0.44\textwidth]
    {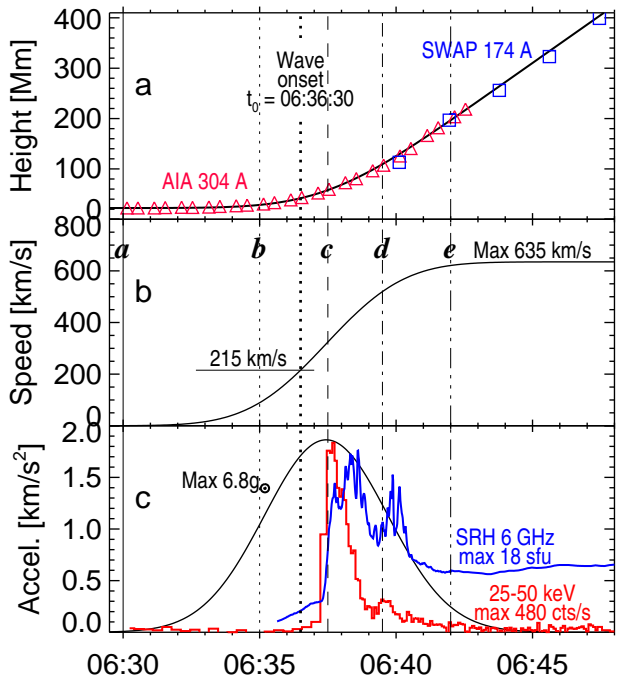}
   }
   \caption{Kinematics of the erupting prominence on 16 March 2016.
a)~Height--time plot measured from the SDO/AIA 304\,\AA\ images
(triangles) and Proba~2/SWAP 174\,\AA\ images (squares). The
analytic curve was fit to the measurements (see the text).
b)~Velocity--time plot. The vertical lines of different styles
denote the times of the images in Figure~\ref{F-2016-03-16_aia304};
its panels are indicated by the bold-italic labels.
c)~Acceleration--time plot. The red curve shows the 25--50\,keV flux
(Fermi/GBM). The blue curve shows the 6\,GHz flux (SRH). }
   \label{F-kinem}
\end{figure}

\cite{Lesovoi2017} measured the kinematics of the erupting
prominence from AIA 304\,\AA\ images. To verify those measurements,
we included the 174\,\AA\ observations in a wider field of view with
the \textit{Sun Watcher using Active Pixel system detector and image
processing} (SWAP: \citealp{Berghmans2006}) onboard the PROBA~2
micro-satellite. Although the rising prominence is barely detectable
in the SWAP images, they allowed us to expand the measured height
interval almost twice. The results are shown in
Figure~\ref{F-kinem}a, where the red triangles represent the
measurements from AIA images, and the blue squares correspond to the
measurements from SWAP images. The refinement of the measurements
did not affect the results considerably.

The height--time dependence in Figure~\ref{F-kinem}a is simple:
The initial speed is close to zero; then the slope (i.e. speed)
monotonically increases and finally becomes nearly constant. The
acceleration determines the curvature of the bend in the
height--time plot; it works within a limited interval and does not
change the sign. The double integration in the transition from the
acceleration to the height--time plot makes the role of a
particular shape of the acceleration pulse negligible. Here we use
a Gaussian acceleration pulse, adjusting its parameters to match
the height--time points measured. The variations in the height,
velocity, and acceleration of the prominence top are calculated in
this way by integration of a smooth analytic function instead of a
problematic differentiation of scattered measured points. The
method of the analytic fit to the measured data proved its
reliability and accuracy in several studies \citep{Gallagher2003,
Sheeley2007, WangZhangShen2009, Alissandrakis2013} and was also
successfully used in the cases, when the kinematics was more
complex (e.g. \citealp{Grechnev2011_I, Grechnev2013_6dec,
Grechnev2016, KuzmenkoGrechnev2017}).

The velocity and acceleration of the prominence top found using
this method are presented in Figures \ref{F-kinem}b and
\ref{F-kinem}c. For comparison, Figure~\ref{F-kinem}c also shows
the temporal profiles of the burst recorded by the SRH at 6\,GHz
and by the \textit{Fermi Gamma-ray Burst Monitor} (GBM:
\citealp{Meegan2009}) in HXR. The maximum velocity acquired by the
prominence top was 635\,km\,s$^{-1}$, much higher than the sound
speed. Hence, plasma ahead of the erupting prominence could not
efficiently flow away which results in the development of a
compression region. The acceleration reached 1.86\,km\,s$^{-2}$,
or 6.8-fold solar gravity acceleration ($g_\odot =
274$\,m\,s$^{-2}$ at the solar surface).

Although the peaks of the HXR burst and acceleration pulse
occurred nearly simultaneously, the prominence started
accelerating at least two minutes earlier than the main sharp rise
of the microwave and HXR bursts. Thus, microwave SRH observations
and HXR data indicate that efficient electron acceleration was
initiated by the prominence eruption. We observed the earlier
development of the eruption process with respect to non-thermal
flare emissions in different events, where a clear lag of order
100\,s was present between the acceleration pulse and flare bursts
\citep{Grechnev2011_I, Grechnev2013_6dec, Grechnev2016}. This
relation does not support an attractive idea of a feedback
relationship between the CME motion and the flare energy release
\citep{Vrsnak2008}.

\subsection{EUV Wave}
 \label{S-EUV_wave}

With a strong acceleration up to $6.8 g_\odot$, the erupting
prominence must have produced a magnetohydrodynamic (MHD) wavelike
disturbance. Its initial propagation velocity is determined by the
local fast-mode speed ($v_\mathrm{fast}$), which is high above an
active region (typically $v_\mathrm{fast} > 1000$\,km\,s$^{-1}$).
Away from the wave origin, the $v_\mathrm{fast}$ in the
environment decreases both upwards and laterally, reaching about
$200$\,km\,s$^{-1}$ above the quiet Sun. When a high-speed
disturbance enters the environment of a considerably lower
$v_\mathrm{fast}$, its profile steepens, and the disturbance
rapidly becomes a shock wave. In this impulsive-piston scenario,
the shock formation is determined mainly by the maximum
acceleration of the eruption and the $v_\mathrm{fast}$ falloff
away from the eruption region and does not depend on the relation
between the eruption speed and the local $v_\mathrm{fast}$ in the
environment \citep{AfanasyevUralovGrechnev2013}.

The disturbance excited by the erupting prominence is visible in
the \url{2016-03-16_AIA171_211.mpg} movie, which presents nearly
simultaneous AIA 171\,\AA\ and 211\,\AA\ images. The diffuse
coronal background was removed from the 171\,\AA\ images on the
left. The 211\,\AA\ running-difference images on the right show
the propagating disturbance. Unlike some other events, no
manifestations of a rim are detectable around the erupting
prominence in either the 211\,\AA\ running differences or the
filtered 171\,\AA\ images, while the latter could reveal the rim
most clearly (see, e.g., \citealp{Grechnev2016}), if it had been
present.

The 211\,\AA\ running-difference images in the movie reveal the
following. At about 06:35, faint structures above the erupting
prominence appeared, which reveals their displacement caused by
the early rise of the prominence (conspicuous due to its black
appearance in the enhanced-contrast images). A bright compression
region above the prominence top appeared at 06:37, when its
velocity reached $300$\,km\,s$^{-1}$, and expanded at 06:38, when
the velocity became $400$\,km\,s$^{-1}$. A fast disturbance
propagated during 06:39--06:42 along transequatorial loops
connecting the parent active region with remote southern regions,
indicating a high Alfv{\'e}n speed in the loops. Then, a
large-scale brightening (EUV wave) is visible that propagates
along the surface and above the limb on the southwest.

To analyze the EUV wave propagation quantitatively, we invoke its
approximate analytic description, which was used in our previous
studies of several events \citep{Grechnev2008, Grechnev2011_I,
Grechnev2011_III, Grechnev2013_6dec, Grechnev2014_II,
Grechnev2015, Grechnev2016, Grechnev2017_III} to follow various
shock-wave signatures such as EUV waves, Type II bursts, and wave
traces ahead of CMEs. This approach uses a power-law density model
\begin{eqnarray}
n(x) = n_0(x/h_0)^{-\delta}
 \label{E-pl_model}
\end{eqnarray}
where $x$ is the distance from the eruption center, $n_0$ is the
density at a distance $h_0 = 100$~Mm, which is close to the scale
height, and the density falloff exponent $\delta$ generally
depends on the wave propagation direction. The development of a
compression region during the eruption before the appearance of
the shock wave strongly disturbs the corona, making standard
coronal density models in the near zone inadequate, while the
corona remains quiet in the far zone. The power-law density model
(\ref{E-pl_model}) describes this situation acceptably: with $x
\approx r-R_\odot$ being the height from the photosphere, $n_0 =
4.1 \times 10^8$~cm$^{-3}$, and $\delta = 2.6$, it is close to the
equatorial Saito model \citep{Saito1970} within $\pm 30\%$ at the
distances exceeding 260\,Mm, providing higher densities at lesser
heights.

A blast-wave-like shock, which spends its energy to sweep up and
extrude the plasma from the volume it occupied previously, has a
power-law kinematics, $x(t) \propto t^{2/(5-\delta)}$ versus time
$t$ \citep{Grechnev2008}. We use this equation in the form
\begin{eqnarray}
x(t) = x_1[(t-t_0)/(t-t_1)]^{2/(5-\delta)},
 \label{E-pl_fit}
\end{eqnarray}
where the starting estimate of the wave onset time, $t_0$, can be
taken equal to the maximum acceleration time, and $x_1$ is the
distance from the eruption center to one of the wave fronts
observed at time $t_1$. Then, we adjust in sequential attempts the
$\delta$ and $t_0$ parameters to reach a best fit of the wave
propagation. The density falloff exponent $\delta$ determines the
curvature of the distance--time plot: with a maximum value $\delta
= 3$ it has a linear shape, and a decrease of $\delta$ increases
the curvature of the plot.

The shape of the global shock-wave front is close to an ellipsoid
\citep{Grechnev2011_III, Grechnev2014_II, Grechnev2017_III,
Kwon2014, Kwon2015, Rouillard2016} with a ratio of the axes not
much different from unity; for simplicity we consider a spheroid,
i.e. ellipsoid of revolution. Its axis corresponds to the
acceleration vector of the eruption. If the large-scale
$v_\mathrm{fast}$ distribution is strongly inhomogeneous (e.g.
because of the presence of a large coronal hole), then the
orientation of the axis gradually displaces toward the region of a
higher $v_\mathrm{fast}$ \citep{Grechnev2011_III,
Grechnev2013_6dec}. The shock front is ``hard'' like an ocean tube
wave, being governed by the global wave expansion and does not
depend on local inhomogeneities in the $v_\mathrm{fast}$
distribution. For this reason, the description of the near-surface
wave propagation with Equation~(\ref{E-pl_fit}) corresponds to an
intermediate value of $\delta_\mathrm{S}$ between zero expected
for a constant density and $\approx 2.6$ typical of the radial
direction (we usually observed $\delta_\mathrm{S} \approx 2.0$ for
EUV waves). The stronger near-surface retardation causes a tilt of
the shock front sometimes observed \citep{Hudson2003,
Warmuth2004_II}. Local inhomogeneities in the $v_\mathrm{fast}$
distribution over the solar surface determine the brightness of
the EUV wave \citep{Grechnev2011_III}, while larger
inhomogeneities affect its propagation velocity and cause its
reflection and refraction (e.g. \citealp{Veronig2008,
Gopalswamy2009, Grechnev2011_I}).

\begin{figure} 
   \centerline{\includegraphics[width=0.48\textwidth]
    {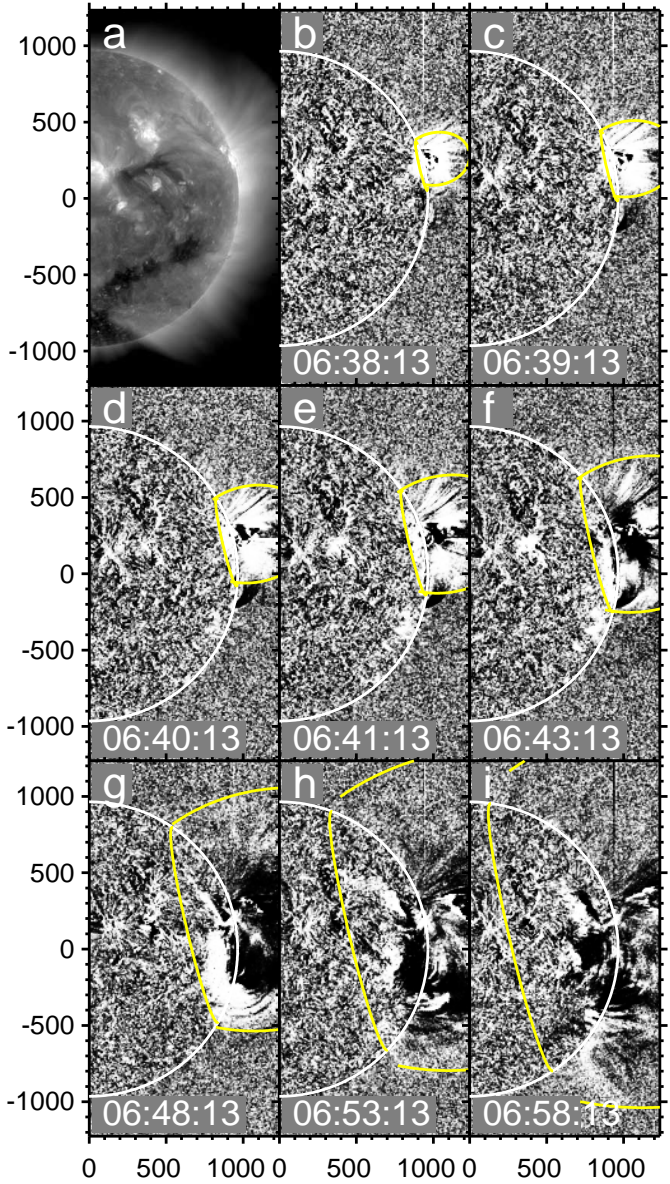}
   }
   \caption{a)~Average of four AIA 211\,\AA\ images on 16 March 2016
from 06:30 to 06:33. b--i)~EUV wave in running-difference AIA
211\,\AA\ images. The white circle denotes the solar limb. The arcs
outline the wave front. The axes indicate the distance from solar
disk center in arcseconds.}
   \label{F-aia_wave}
\end{figure}

Keeping in mind these circumstances, we calculated the global
shock-wave fronts and their surface skirt (EUV wave). They are
shown in Figure~\ref{F-aia_wave}b--i and the
\url{2016-03-16_AIA211_wave.mpg} movie on top of the AIA~211\,\AA\
running differences. Figure~\ref{F-aia_wave}a presents an averaged
pre-event AIA~211\,\AA\ image, which shows active regions (green
in the movie) and coronal holes (blue in the movie). The elliptic
arcs on the surface are small circles parallel to the equator of
the sphere, whose pole coincides with the eruption site. The
distances are measured from the pole to the small circles along
the great circle.

Figure~\ref{F-aia_wave} and the movie reveal a complex character
of the EUV wave. From 06:37 to 06:53, the calculated ellipses
bound its outermost signatures in both hemispheres, except for the
mentioned southwards fastest disturbance on the west above the
limb. After 06:53, the EUV wave is conspicuous southwest from the
extended southern coronal hole, while large-scale inhomogeneities
complicate and hamper its propagation farther in the northern
hemisphere. Overall, while the calculated ellipses represent, on
average, the global expansion of the wave dome above the limb and
its surface trail, the presence of active regions and coronal
holes governs the propagation and appearance of the EUV wave
according to the associated inhomogeneities in the
$v_\mathrm{fast}$ distribution over the solar surface. Their
influence corresponds to the expectations for a mast-mode wave.

\begin{figure} 
   \centerline{\includegraphics[width=0.48\textwidth]
    {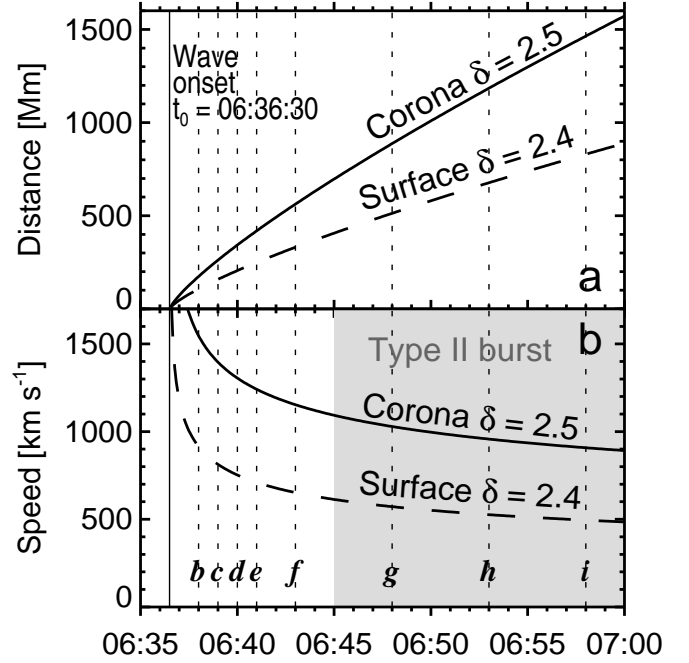}
   }
   \caption{Distance--time (a) and velocity--time (b) plots of the EUV wave.
The wave propagation in the radial direction (up) is represented
by the solid curve, and the dashed curve represents its surface
trail. The vertical solid line denotes the wave onset time. The
vertical dotted lines denote the times of the images in
Figure~\ref{F-aia_wave} whose panels are indicated by the
bold-italic letters. The shading in panel b denotes the
observation interval of the Type~II burst (continued afterward).}
   \label{F-aia_wave_kinem}
\end{figure}

Figure~\ref{F-aia_wave_kinem} presents the kinematics used to
outline the wave signatures in Figure~\ref{F-aia_wave} and the
movie. The wave onset time was refined to fit the EUV wave
propagation, $t_0 = $06:36:30 (the vertical thick-dotted line in
Figure~\ref{F-kinem}). The density falloff exponents for the
radial direction $\delta_\mathrm{C} = 2.5$ and for the
near-surface propagation $\delta_\mathrm{S} = 2.4$ almost coincide
in this case.

The EUV wave velocity in Figure~\ref{F-aia_wave_kinem}b
monotonically decreased by 80\% within an interval shown in
Figure~\ref{F-aia_wave}. This behavior with a strong deceleration
is consistent with a pioneering result of \cite{Warmuth2001} and
several later studies, but is not exhibited by all EUV transients
(e.g. \citealp{Warmuth2004_I, Warmuth2004_II, Warmuth2005,
Muhr2011, Muhr2014, Nitta2013_waves, Long2017}; see
\citealp{Warmuth2015} for a review). In our previous case studies,
we observed exactly this behavior for shock-associated EUV waves.
On the other hand, if the EUV wave properties had been studied
solely from signatures in the images, especially by means of an
automated detection algorithm, then understanding its kinematics
would be difficult.

\subsection{Type II Burst}
 \label{S-type_II}

While the EUV wave reveals a fast-mode disturbance, which was most
likely super-Alfv{\'e}nic, its shock-wave regime is not obvious. A
commonly accepted evidence of a shock wave is a Type~II radio
burst. An important property of Type~II bursts is their
narrow-band emission. To ensure it, the source should be compact;
otherwise, a large shock front crossing a wide range of plasma
densities could only produce a drifting continuum
\citep{KnockCairns2005}. An appropriate source of a Type~II
emission is a distinct narrow structure, i.e. coronal streamer
\citep{Uralova1994, Reiner2003} that was confirmed in imaging
meter-wave observations of Type~II sources \citep{Feng2013,
Chen2014, Du2014, Lv2017}. A Type~II burst can be emitted from a
remote streamer crossed by a flank of a quasi-perpendicular or
oblique shock or from a streamer located above the eruption region
crossed by the front of a quasi-parallel shock. The former case
probably corresponds to a typical situation, and the infrequent
latter case is characterized by a considerably faster drift
\citep{Grechnev2014_II, Grechnev2016}. In either case, the shock
crossing the streamer deforms its current sheet that produces a
flare-like process running along the streamer together with the
intersection point. This scenario has shed light on various
structural properties of Type~II bursts \citep{Grechnev2011_I,
Grechnev2014_II, Grechnev2015, Grechnev2016}.

\begin{figure} 
   \centerline{\includegraphics[width=0.48\textwidth]
    {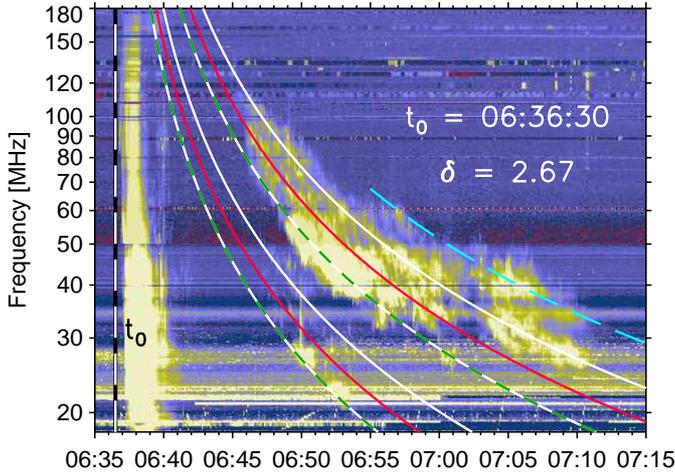}
   }
   \caption{Dynamic spectrum of the metric radio burst composed from
the Learmonth and Culgoora data. The vertical dashed line denotes
the wave onset time $t_0 = $\,06:36:30. The curves of different
line styles and colors outline different bands in the Type~II
structure; the paired curves outline the fundamental and harmonic
emissions. All of the curves correspond to the same $t_0$ and
density falloff exponent $\delta = 2.67$, suggesting a single
shock front crossing a few emitting structures.}
   \label{F-dyn_spec}
\end{figure}

Figure~\ref{F-dyn_spec} shows a dynamic spectrum combined from the
Learmonth and Culgoora spectrographs. The spectrum presents a
strong Type~V burst co-temporal with the main burst in HXR and
microwaves in Figure~\ref{F-kinem}c followed by a faint Type~III
burst at 06:40 corresponding to a minor burst. At 06:46, a Type~II
burst with a complex structure started. Its fundamental-emission
band was strongly suppressed, while the harmonic emission
consisted of at least three indistinct lanes. A fine Type~III-like
structure of the lanes is detectable suggesting acceleration of
electrons in the running flare-like process.

To analyze the frequency--time drift of the Type~II burst, we use
the approach described in the preceding section. We choose a
reference point of a Type~II band on the dynamic spectrum at time
$t_1$ with a frequency $f_1$, convert the frequency into the
density $n_1$ assuming the fundamental emission at the plasma
frequency $f_\mathrm{P} = 9 \times 10^4 n^{1/2}$ or its second
harmonic $2 f_\mathrm{P}$, and then convert $n_1$ into the
distance $x_1$ using the power-law density model
(\ref{E-pl_model}). Taking starting estimates for $t_0$ and
$\delta$, we calculate the trajectory using
Equation~(\ref{E-pl_fit}), convert it to the frequency and plot on
top of the dynamic spectrum. The values of $t_0$ and $\delta$ are
optimized in sequential attempts to reach the best fit of the
trajectory to bright Type~II signatures (see
\citealp{Grechnev2014_II, Grechnev2017_III} for details). If a
Type~II band is clearly defined, then two reference points can be
chosen. The \url{type_II_fit.mpg} movie presents the adjustment of
the Type~II trajectory using this example. Here the only variable
is $\delta$, which governs the curvature of the trajectory, and
its optimal value $\delta = 2.67$ determines $t_0 =$\,06:36:30,
the same as for the EUV wave. The difference between the $\delta =
2.67$ and $\delta_\mathrm{C} = 2.50$ for the coronal wave
(Figure~\ref{F-aia_wave_kinem}) can be due to different
directions.

With $t_0$ and $\delta$ estimated for a single harmonic band, the
trajectories for different bands at both harmonics were calculated
by referring to different $f_1$ at the same $t_1$ and plotted in
Figure~\ref{F-dyn_spec} with different line styles and colors
(same for each harmonically related pair). An extra band with the
same $t_0$ and $\delta$ appeared at 06:55:00. The coincidence of
the wave onset times and even the density falloffs for all of the
bands indicates their common origin related to the same shock
front.

The structure of the Type~II burst does not resemble the
band-splitting, and this effect conventionally interpreted by the
emission upstream and downstream of the shock front cannot account
for more than two bands. It is also difficult to relate this
structure to a single bow-shock-associated source ahead of the CME
nose, which can only produce a single or split harmonic pair of
bands. Instead, the presence of several pairs of bands points at a
corresponding number of compact sources not much different from
each other. Most likely, they were located at the flanks of the
coronal wave and none ahead of the CME nose because of their
similar drift rates with the same $\delta$. This assumption is
supported by the strong absorption of the fundamental emission
along the line of sight either in a long column of the corona in
front of the Type~II sources above the west limb, or a dense
structure such as the base of the streamer belt, or both.

The appearance of the EUV wave in Figure~\ref{F-aia_wave} and the
\url{2016-03-16_AIA211_wave.mpg} movie is not different before the
start of the Type~II burst (06:45:00) and after it. The wave
velocity in Figure~\ref{F-aia_wave_kinem}b monotonically
decreased, being in the first panels of Figure~\ref{F-aia_wave}
most likely higher than the ambient fast-mode speed both along the
surface and in the radial direction. The Type~II burst started
when the wave considerably decelerated (shading in
Figure~\ref{F-aia_wave_kinem}b). All of these facts indicate that
the lag of the Type~II burst behind the wave onset time is
determined by the distance required for the shock front, which
already exists, to propagate until the encounter with a streamer,
which can produce the Type~II emission, and does not depend on the
relation between the velocity of the wave or ejecta and the
ambient fast-mode speed. \cite{Long2017} found the delay of a
Type~II burst relative to the EUV wave onset to be typical.

In summary, both the EUV wave and Type~II burst point to the same
wave onset time at 06:36:30. The velocity of the prominence top,
which excited the wave, was 215\,km\,s$^{-1}$ at that time (the
thick dotted line in Figure~\ref{F-kinem}b). It should be noted
that Equation~(\ref{E-pl_fit}) used in our measurements was
obtained for a spherical blast wave expanding from a point-like
source \citep{Grechnev2008}. A real wave exciter can be spatially
extended, which might shift the actual wave onset time. In the
radial direction corresponding to the eruption, the wave
represented by the solid curve in Figure~\ref{F-aia_wave_kinem}
travels, e.g., 20\,Mm in 6\,s and 50\,Mm in 20\,s. Even with the
largest time shift the velocity of the prominence top in
Figure~\ref{F-kinem}b did not exceed 300\,km\,s$^{-1}$, being
certainly sub-Alfv{\'e}nic. On the other hand, the wave started
close to the maximum acceleration time in Figure~\ref{F-kinem}c
that occurs in the impulsive-piston shock excitation scenario.

\subsection{White-Light Transient}
 \label{S-CME}

\begin{figure} 
   \centerline{\includegraphics[width=0.48\textwidth]
    {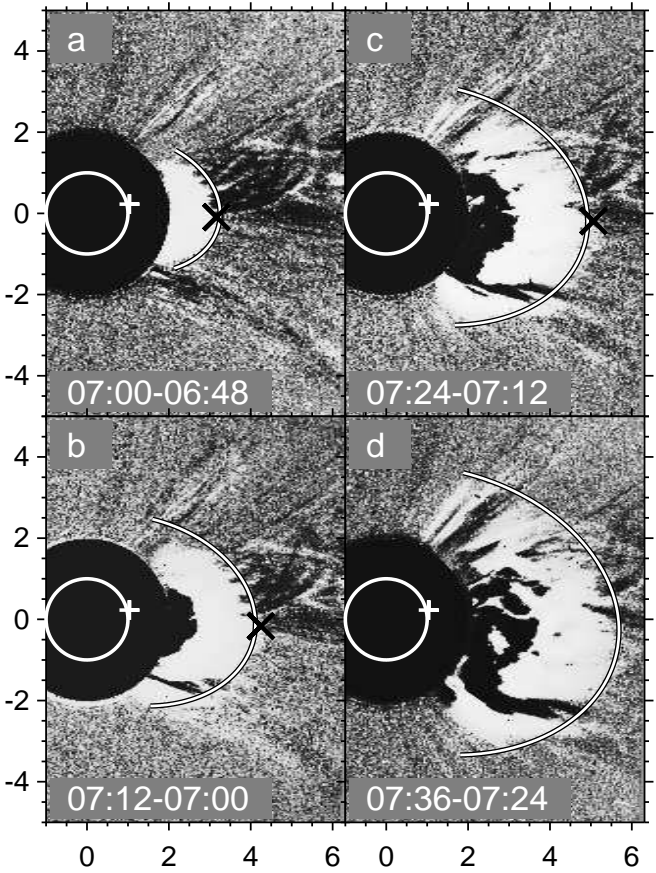}
   }
   \caption{Wave traces on 16 March 2016 in LASCO-C2 images (running differences).
The thick white circle denotes the solar limb. The small white
crosses denote the eruption center. The larger slanted black crosses
in panels a--c denote the measurements in the CME catalog. The arcs
outline the wave front. The axes indicate the distance from solar
disk center in solar radii.}
   \label{F-2016-03-16_c2_wave}
\end{figure}

The eruption produced a decelerating CME. According to the online
CME catalog (\url{https://cdaw.gsfc.nasa.gov/CME_list/}:
\citealp{Yashiro2004}), it had a central position angle of
$265^{\circ}$, an average speed of 592\,km\,s$^{-1}$, and
acceleration of $-22.4$\,m\,s$^{-2}$.
Figure~\ref{F-2016-03-16_c2_wave} presents the wave traces in
contrasted LASCO-C2 running-difference images. The radii of the
white-on-black arcs were calculated from the decelerating wave
kinematics in Figure~\ref{F-aia_wave_kinem}a with the same
$t_0=$\,06:36:30 and $\delta = 2.5$. The arcs match most of the
wave traces, which are manifested in the partial halo enveloping
the CME body and deflections of the coronal rays. The arcs are
close to the measurements in the CME catalog denoted by the black
slanted crosses.

\begin{figure} 
   \centerline{\includegraphics[width=0.48\textwidth]
    {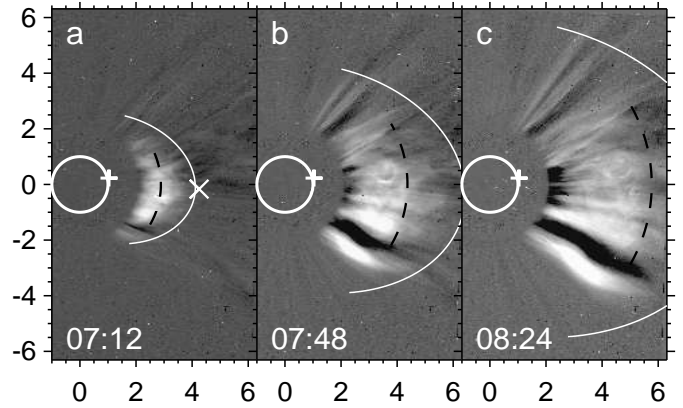}
   }
   \caption{The CME on 16 March 2016 in LASCO-C2 images (fixed-base ratios).
The thick white circle denotes the solar limb. The small crosses
denote the eruption center. The slanted cross in panel a denotes the
measurement in the CME catalog. The white solid arcs outline the
wave front, and the black-dashed arcs outline the flux-rope-like
structure. The axes indicate the distance from solar disk center in
solar radii.}
   \label{F-2016-03-16_c2}
\end{figure}

Figure~\ref{F-2016-03-16_c2} shows the CME structure in
non-subtracted C2 images. The white arcs correspond to wave
traces. Neither the frontal structure nor cavity are pronounced.
The black-dashed arcs outline the main part of the CME body (core)
with a helical structure inherited from the erupted prominence. It
seems to be more complex than one expects for a perfect flux-rope
structure.

\begin{figure} 
   \centerline{\includegraphics[width=0.48\textwidth]
    {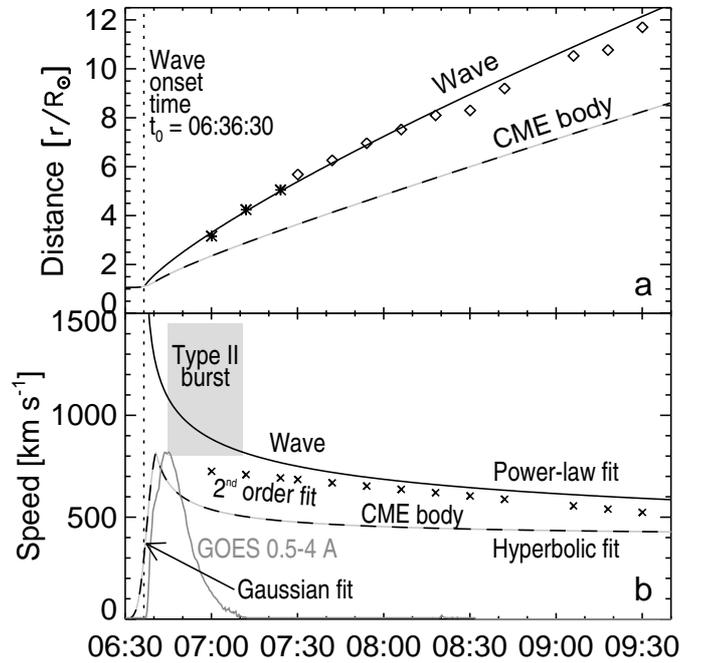}
   }
   \caption{Overall kinematical plots of the wave signatures (solid)
and CME body (dashed): a)~heliocentric distances versus time,
b)~velocity--time plots. The symbols represent the measurements in
the CME catalog. The gray curve in panel b is the GOES
0.5--4\,\AA\ flux scaled to match the plot of the CME body. The
shading in panel {b} shows the interval when the Type~II burst was
observed.}
   \label{F-cme_kinem}
\end{figure}

Figure~\ref{F-cme_kinem} presents the kinematical plots for the
wave (solid) and CME body (dashed) along with the measurements
from the CME catalog (symbols). The way to obtain the wave
kinematics has been discussed in detail. It is more complex to
infer the kinematics of the CME body, which is determined by
different processes at different stages of its development.

The kinematics of the erupting prominence governed by an MHD
instability was measured in Section~\ref{S-prom_eruption} using
the fit with a Gaussian acceleration pulse (Figure~\ref{F-kinem}).
When the instability expires, the CME expands for some time freely
and self-similarly \citep{CremadesBothmer2004}. Eventually, the
CME kinematics should be determined by the aerodynamic drag from
solar wind \citep{Chen1989, Chen1996, VrsnakGopal2002}, whose
dominance is expected beyond $15\,\mathrm{R}_\odot$
\citep{Vrsnak2006, Temmer2011}. As \cite{KuzmenkoGrechnev2017}
showed, exceptions do occur, and nevertheless the CME expands
nearly self-similarly at moderate distances from the Sun.

The self-similar character of the CME expansion is determined by
the fact that the magnetic propelling and retarding forces, plasma
pressure and gravity decrease after the termination of the initial
instability by the same factor inversely proportional to the
distance from the eruption center squared (except for the drag).
The theory of self-similar expansion of solar CMEs was initially
developed by \cite{Low1982}. A description of a self-similar
expansion convenient in the analysis of observations was proposed
by \cite{UralovGrechnevHudson2005}. From their formulas, the
instantaneous velocity $v$ can be related to the distance $R$ from
the expansion center \citep{Grechnev2008}:
\begin{eqnarray}
v^2 = v_0^2+\left(v_\infty^2 - v_0^2\right)\left({1-R_0/R}\right),
 \label{E-self-sim_vel}
\end{eqnarray}
where $R_0$ is the initial position of the CME and $v =
\mathrm{d}R/\mathrm{d}t$ and $v_0$ and $v_\infty$ are its initial
velocity and the asymptotic final velocity in the self-similar
expansion stage. With a simple form,
Equation~(\ref{E-self-sim_vel}) cannot be integrated explicitly;
the formulas for the time versus distance dependence are
cumbersome. They can be found in \cite{Grechnev2014_II}. The
properties of the self-similar plots correspond to those of
hyperbolic functions. Acceleration in the self-similar regime
cannot increase by the absolute value and therefore this approach
does not apply to the CME's initial lift-off during the impulsive
acceleration stage.

We concatenated the kinematics of the erupting prominence fitted
with a Gaussian acceleration (Figure~\ref{F-kinem}) with the
self-similar kinematics of the CME. The rising prominence forces
to expand closed coronal structures above it, which are expected
to be ahead but were not observed. To take account of their
presence in LASCO images, the prominence velocity was increased by
40\%. The resulting velocity--time plot for the CME body is
presented in Figure~\ref{F-cme_kinem}b by the dashed curve. Its
integration provided the distance-time plot in
Figure~\ref{F-cme_kinem}a used to calculate the radii of the
black-dashed arcs outlining the CME body in
Figure~\ref{F-2016-03-16_c2}. The
\url{2016-03-16_C2_rope_wave.mpg} movie shows the CME body and
wave in the images, whose field of view is scaled according to the
measured kinematics to fix the visible size of the transient.

\cite{Zhang2001} established similarity between the CME velocity
variations and the rise phase of the GOES SXR flux and found
indications of similarity between the CME acceleration and the HXR
burst confirmed by \cite{Temmer2008}. The similarity between the
HXR and the derivative of the SXR flux is really expected due to
the Neupert effect \citep{Neupert1968}. A case study by
\cite{Grechnev2016} demonstrated a close correspondence between
the kinematics of an erupting structure and X-ray emissions, which
were delayed by about 2 minutes that resembles the situation in
this event. There is the similarity indeed between the rising
parts of the CME velocity plot and the GOES 0.5--4\,\AA\ flux
(gray in Figure~\ref{F-cme_kinem}b), which lags behind the
velocity by 140\,s.

The self-similar plots resemble the CME kinematics expected for a
drag-dominated situation, whereas the responsible forces are quite
different (the similarity is also possible for
gradually-accelerating slow CMEs). For this reason, if a
drag-based model acceptably describes the CME kinematics, then
this result does not guarantee the importance of the drag.

The measurements in the CME catalog are carried out for the
fastest feature of a transient, being therefore most likely
related to a wave ahead of the CME body, if it is present.
Figure~\ref{F-cme_kinem}a confirms the agreement between these
measurements and our curve. To find the velocity of a transient,
the linear and second-order fit are used in the CME catalog. The
latter is presented in Figure~\ref{F-cme_kinem}b by the slanted
crosses, whose difference from our power-law fit is mostly not
large. The difference increases at shorter distances that results
in a strong underestimation by the second-order fit of the wave
velocity during its initial evolution hidden by the occulting disk
of LASCO-C2.

The interval when the Type~II burst was observed is denoted in
Figure~\ref{F-cme_kinem}b by the gray shading. The Type~II burst
ceased by 07:11, when the wave velocity decreased to about
800\,km\,s$^{-1}$, and did not extend into the frequency range
below 14\,MHz. These circumstances indicate that the decelerating
shock decayed at about this time into a weak disturbance. The
maximum heliocentric distance at that time was
$4.2\,\mathrm{R}_\odot$ for the wave front and
$2.8\,\mathrm{R}_\odot$ for the CME body, whose velocity was
500\,km\,s$^{-1}$. The shock wave had not changed to the bow-shock
regime, because the trailing CME body was sub-Alfv{\'e}nic.

\subsection{Implication to the Near-Earth Proton Enhancement}

\begin{figure} 
   \centerline{\includegraphics[width=0.48\textwidth]
    {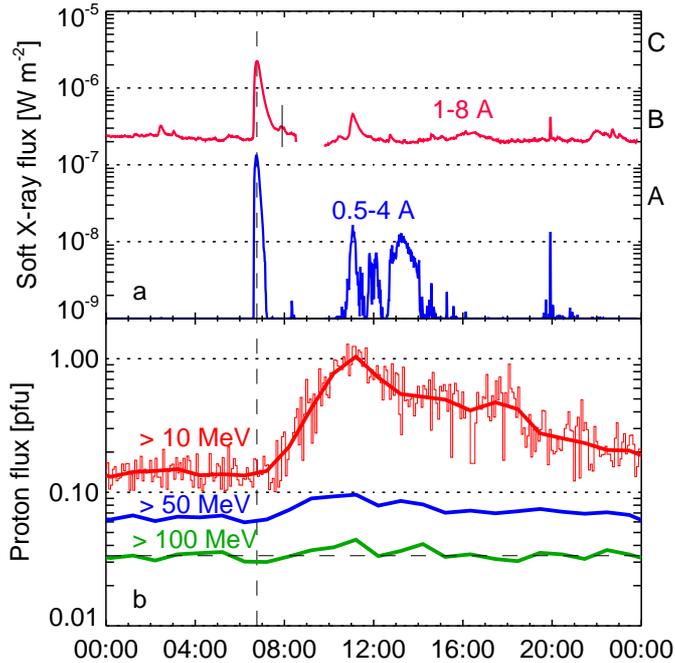}
   }
   \caption{GOES plots of SXR fluxes (a) and integral
proton channels (b) recorded on 16 March 2016. The histogram-like
thin line in panel b presents the original 5-minute data on $>
10$\,MeV protons. The thick lines present the proton fluxes summed
over 1~hour. The vertical dashed line marks the peak time of the
SXR flux. The horizontal dashed line shows the background level in
the $> 100$\,MeV proton channel averaged over the preceding and
next days.}
   \label{F-xray_protons}
\end{figure}

The SXR emission of this eruptive flare up to C2.2 level had an
impulsive time profile with a duration of 23 minutes
(Figure~\ref{F-xray_protons}a). At about the time of the event, a
weak near-Earth proton enhancement started
(Figure~\ref{F-xray_protons}b). The proton flux reached about
1\,pfu in the $> 10$\,MeV integral channel, was detectable in the
averaged $> 50$\,MeV channel, and possible in the $> 100$\,MeV
channel, exceeding the $3 \sigma$ level above the background
around 11:00. Figure~\ref{F-xray_protons}a also reveals a minor
secondary SXR enhancement during 07:45--08:05 marked on the
1--8\,\AA\ plot by a thin vertical bar. A group of metric
Type~IIIs around 08:00 extending to lower frequencies in the
Wind/WAVES spectrum corresponds to this minor event, while neither
SOHO/LASCO nor STEREO-A/COR1 show any additional CME. The proton
event already started at that time and was therefore caused by the
eruptive C2.2 event in AR\,12522 observed by the SRH, while the
minor event around 08:00 was unlikely important.

This impulsive flare accompanied by a modest microwave burst of
18~sfu seems to be too weak to produce the proton event; the most
probable candidate for its source is the shock wave. It appeared
during the flare rise, being able to accelerate protons
considerably earlier than usually assumed, and decayed soon,
having not changed to the regime of the CME-driven bow shock.
These circumstances show that a widely accepted view relating
solar energetic particles with CME-driven shocks, which develop at
considerable heights, needs refinement.

\section{Discussion}
\label{S-discussion}

\subsection{Summary on the Eruptions Observed with the SRH}
 \label{S-summary_eruptions}

Being not able to resolve the spatial structure of eruption
regions, the SRH detects the occurrence of many eruptions, whose
energy and spatial size can be very small, and locates their
positions on the Sun. The eruptions presented here were revealed
in one of three ways: i)~from microwave depressions shown in
Section~\ref{S-neg_bursts}, ii)~by direct SRH observations of the
eruptions, as was the case on 1 May 2017 in
Section~\ref{S-eruption_may_1}, and iii)~by the examination of the
eruptive flare observed by the SRH on 16 March 2016
(Section~\ref{S-march16}). In all of these cases, the SRH provides
the pointing to the events, which are analyzed using data acquired
by a number of different instruments. This is a usual way to study
complex solar events.

The microwave depressions shown in Section~\ref{S-neg_bursts} as
well as the negative bursts on 9 August 2016 in AR\,12574
(N04\,E59) presented by \cite{Lesovoi2017} occurred not far from
the limb. The intensity depressions were accompanied by changes in
the polarization that indicates the screening of microwave
sources, which had a considerable polarization, i.e. gyromagnetic
sources. In all of these cases, the screening was caused by
low-temperature jets, which occurred near sunspots indeed. Thus,
the jet-like eruptions responsible for the depressions most likely
screened polarized sunspot-associated sources. Because the
orientations of the jets are not much different from the radial
direction, the screening phenomena are favored by the location of
the eruptions close to the limb. Deviations in the Stokes~V
correlation plots indicate such events, as a cursory analysis of
different depressions observed by the SRH confirms.

The events considered in Sections \ref{S-neg_bursts} and
\ref{S-eruption_may_1} were associated with jet-like eruptions of
different size, where the low-temperature erupted plasma rose and
gradually crossed in front of microwave sources, absorbing their
emission. The screening caused long-lasting depressions of the
microwave emission and changes in its polarization.

As noted in Section~\ref{S-neg_bursts}, multi-frequency
observations of microwave depressions provide the basis for plasma
diagnostics in erupting structures. Modeling the spectrum of the
absorption depths observed at a few frequencies from 1 to 10\,GHz
allowed estimating parameters of the erupted material responsible
for several negative bursts even without images
\citep{Grechnev2008, Grechnev2013neg, KuzmenkoGrechnevUralov2009,
KuzmenkoGrechnev2017}. These studies used a flat-layered model of
a relatively large absorbing cloud of given height, dimensions,
temperature, and density, with a possible stable compact microwave
source covered. The estimated area screened reached 2--10\,\% of
the solar disk. This approach can also be used to analyze from the
SRH data large-scale absorption phenomena, when they would be
observed. A narrower SRH frequency range of 4--8\,GHz relative to
these studies might result in increased uncertainties.

Plasma diagnostics for small eruptions shown in
Section~\ref{S-neg_bursts} is more complex. The fractions of the
solar disk covered by the jets in 304\,\AA\ were about 0.12\,\% on
9 September 2017, 0.03\,\% on 3 August 2017, and 0.05\,\% on 1 May
2017. The small width of the screen becomes comparable with the
size of the microwave source behind it. Here it is necessary to
consider additionally the overlap between the narrow jet and a
microwave source and to untangle the variations in the opacity of
the jet and the changes in the brightness and spectrum of the
flaring source. These issues should be addressed in future
studies.

Most of the events presented here were associated with jet-like
eruptions. A realistic explanation of jets was proposed by
\cite{Filippov2009} and \cite{Meshalkina2009} based on
three-dimensional magnetic configurations containing coronal null
points. Such configurations appear above photospheric magnetic
islands surrounded by opposite-polarity regions and resemble an
inverted funnel or helmet. If a small flux rope erupts inside the
funnel, then its magnetic structure cannot survive when passing at a
null point \citep{Uralov2014}, and released plasma flows out as a
jet. Eruptions in such configurations are characterized by circular
ribbons and impulsive temporal profiles \citep{Masson2009,
Meshalkina2009}. Magnetic islands inside opposite-polarity regions
occur very often, and inverted funnels (helmets) are also expected
to be quite common configurations. For example, similar
configurations are conjectured in Figures
\ref{F-2017-05-01_aia211_304}b and \ref{F-2017-05-01_aia211_304}d.
The roles of such configurations deserve further attention to be
paid elsewhere.

\subsection{Initiation of an Eruption and Development of a Flux Rope}
 \label{S-flux-rope}

All of the eruptions considered here started developing from below
at small heights in the corona. This circumstance is obvious for
small eruptions presented in Section~\ref{S-neg_bursts} and a
larger event on 1 May 2017 shown in
Section~\ref{S-eruption_may_1}. The situation was also similar in
the CME-related 16 March 2016 event. We consider this event in
more detail.

The main active structure observed in this event was the eruptive
prominence. Its motion started before the HXR and microwave
bursts, and the flare ribbons developed later. The chain of events
resembles the scenario by \cite{Hirayama1974}, in which an MHD
instability of an electric current in the prominence drives its
lift-off, which stretches associated magnetic fields, forming the
current sheet, in which the flare reconnection occurs, and a shock
wave is generated ahead of the erupting prominence. None of the
SDO/AIA 304\,\AA, 171\,\AA, or 211\,\AA\ channels, capable of
detecting non-flaring structures, reveal within the AIA field of
view any larger feature embracing the prominence that could govern
its eruption. The behavior of the erupting prominence in
Figure~\ref{F-2016-03-16_aia304} and the
\url{2016-03-16_AIA193_304_SRH_Fermi.mpg} movie indicates its own
twist instability rather than a reflection of external processes
in a larger structure, whose presence is often presumed.

The structure of a pre-eruptive prominence is considerably different
from a flux rope, which is rooted to the surface by two ends only.
The presence of numerous barbs indicates a multitude of
flux-rope-like segments arranged along the magnetic neutral line,
each of which is connected to the surface by its ends. A presumable
scenario, in which reconnection forms a single flux rope from a
multitude of sheared field lines with the appearance of flare loops
was theoretically described by \cite{InhesterBirnHesse1992} and
\cite{LongcopeBeveridge2007} and got a quantitative support in
observational studies (e.g. \citealp{Qiu2007, Miklenic2009}).

The MHD instability, which governs the initiation and development
of the prominence eruption, is presumably driven by an electric
current. In pre-eruptive force-free conditions, $\nabla \times
\textbf{\textit{B}} = \alpha\textbf{\textit{B}}$; the density of
the electric current is proportional to the magnetic field
strength in a prominence. The field strength in its environment
above an active region steeply falls off, as the height increases
(e.g. \citealp{Gary2001, Mann2003}). Therefore, the magnetic field
and electric current in a prominence are typically stronger near
the solar surface than at larger heights. To produce the
acceleration with a half-height duration of 5 minutes observed for
the erupting prominence in Figure~\ref{F-kinem}, the
characteristic Alfv{\'e}n time in the responsible processes should
be much shorter. This would not be possible if the eruption had
been governed by a large-scale structure with a weaker magnetic
field and longer Alfv{\'e}n time.

The reconnection process detaches the barbs under the prominence,
transforming its structure into the helical structure of the
developing flux rope. When its central part is nearly formed, it
becomes convex, and the torus instability develops.
Figure~\ref{F-2016-03-16_aia304}c presents an episode of this stage
corresponding to the maximum acceleration measured. Then, the twist
instability develops in Figures \ref{F-2016-03-16_aia304}d and
\ref{F-2016-03-16_aia304}e, which is often observed, but does not
seem to be a necessary phase of the eruption process.

The flux-rope formation is unlikely to occur perfectly and terminate
completely in the course of the prominence eruption. Some of the
pre-eruptive segments could not reconnect. The flux-rope-like
structures actually observed (e.g. \citealp{Cheng2013,
Grechnev2016}) resemble twisted bundles of loops rather than a
perfect croissant-shaped structure. \cite{KuzmenkoGrechnev2017}
revealed indications of an ongoing flux-rope formation from twisted
core structures during the CME expansion. The structure of the CME
body in Figure~\ref{F-2016-03-16_c2} observed on 16 March 2016 also
seems to be more complex than an expected croissant-like flux rope
in the CME cavity.

These circumstances indicate that a flux rope forms in the course
of a time-extended process. The eruption observed in the extreme
ultraviolet is its most impulsive, powerful stage, when a future
CME structure develops, while its components have not yet
constituted the whole. This fact is essential to determine the
actual shock-wave excitation scenario.

\subsection{Shock Excitation Scenarios}
 \label{S-scenarios}

The impulsive-piston shock-wave excitation scenario revealed in
Section~\ref{S-march16} is not exceptional. The main conditions
necessary to realize this scenario are i)~more or less impulsive
acceleration of an eruptive structure, and ii)~pronounced falloff
of the fast-mode speed away from the eruption region. These
conditions are typical of many events, irrespective of the flare
size, and even in cases where non-thermal bursts are not observed
in HXR or microwaves. An abrupt eruption is only required, while
the presence of a CME is not necessary.

On the other hand, the impulsive-piston scenario is not expected
for gradually accelerating CMEs initiated by the eruptions of
large quiescent prominences away from active regions. It is also
not expected for confined flares independent of their size, that
are not associated with expansion of any structures. Such rare
flares sometimes occur (e.g. \citealp{Thalmann2015}; a few major
confined flares also occurred in September 2005).

While the shock-wave excitation scenarios have been known for
several decades, observations until recently did not allow
identifying which one was responsible for the appearance of
coronal shock waves (see \citealp{VrsnakCliver2008} for a review).
The search for their origins has been focused on the
``impulsive-piston shock excitation by a flare pressure pulse
versus the bow-shock excitation by the outer surface of a
super-Alfv{\'en}ic CME'' alternative. A rather obvious scenario
outlined in Section~\ref{S-EUV_wave} has been escaping attention,
possibly because the flux ropes are assumed pre-existing when the
eruptions develop.

Having adopted the ``flare versus CME'' alternative, one is
constrained by its framework and comes to a conclusion about the
flare-related shock origin, if its exciter exhibits impulsive
properties (e.g. in the case of Moreton waves), or if mismatch
between the estimated speeds of the shock and CME is conspicuous,
especially if a CME is absent. However, the role of the flare
pressure in the shock-wave excitation is unlikely
\citep{Grechnev2011_I, Grechnev2015} for the following reasons.

\begin{enumerate}

\item
 The plasma density and temperature in flare loops are
manifested in their SXR emission. It is gradual in nature and
resembles the indefinite integral of the HXR burst (the Neupert
effect: \citealp{Neupert1968}). On the other hand, the HXR burst
roughly corresponds to a sharp acceleration of an eruption, which
produces a strong MHD disturbance, while the plasma pressure in
flare loops increases gradually.

\item

The plasma pressure in flare loops cannot considerably exceed the
magnetic pressure, being compensated by the dynamic pressure of
the reconnection outflow. Even if the plasma pressure in a loop
becomes comparable with the magnetic pressure ($\beta \approx 1$),
the effect is as small as an increase in each of its three
dimensions by a factor of $(\beta + 1)^{1/4}$ (see
\citealp{Grechnev2006} for details). The increase in the volume of
flare loops is basically insufficient to produce an appreciable
MHD disturbance outward.

\end{enumerate}

These considerations were verified in case studies of a few
events, in which the presence of shock waves was undoubted and
their onset times were estimated with certainty
\citep{Grechnev2011_I, Grechnev2015}. The plasma pressure in flare
loops estimated from SXR GOES fluxes steadily rose, when the waves
were excited near the peak time of the impulsive acceleration of
an eruption. The size of the SXR-emitting regions in RHESSI images
did not change around the wave onset time. In some events, the
wave onset time clearly corresponded to the early rise of an HXR
or microwave burst, when the chromospheric evaporation responsible
for the plasma pressure in flare loops just started
\citep{Grechnev2013_6dec, Grechnev2014_II, Grechnev2015,
Grechnev2016}. The same situation is seen in Figure~\ref{F-kinem}
in the 16 March 2016 event. The conclusions drawn from the case
studies are supported by the statistical independence of the EUV
wave occurrence on the flare size \citep{Long2017}.

While the relation between the velocity of an eruption and the
ambient fast-mode speed is not important for the initial
impulsive-piston excitation of a shock wave, it is crucial for its
later evolution. A decelerating shock wave is supplied by the
energy from the trailing ``piston'', whose role at larger
distances really plays the outer surface of the CME body. If it is
fast, then the shock wave changes into the bow-shock regime. If
the CME is slow, as was the case in the 16 March 2016 event, then
the shock decays into a weak disturbance. This occurs most rapidly
in confined eruptions without CMEs (but not confined flares). Very
rare events of this kind are known indeed, in which EUV waves or
Type~II bursts, or both were observed (e.g.
\citealp{Shanmugaraju2006, Magdalenic2012, Nitta2014,
Grechnev2014_II, Eselevich2017}). Thus, the fact that the vast
majority of EUV waves are associated with CMEs (e.g.
\citealp{Biesecker2002, Long2017}) does not guarantee that every
shock wave has an associated CME.

The studies of shock-wave histories are facing heavy observational
difficulties. Eruptive structures rapidly acquire high velocities
and dramatically lose brightness. Wave signatures possess strong
initial deceleration, which is most conspicuous in the first few
minutes of their propagation, as Figure~\ref{F-aia_wave_kinem}b
exemplifies. At that time, the measurements of the wave propagation
and even its detection are hampered by a strong flare emission,
while the imaging rate and dynamic range of telescopes are limited.
In addition, different objects appear similar to shock-related EUV
waves --- for example, rising CME structures and quasi-stationary
compression regions at their bases \citep{ZhukovAuchere2004,
ChenFangShibata2005, Grechnev2011_III, Warmuth2015}. Finally, a
shock wave excited by a sharply erupting structure has a kinematics
similar to what is expected for a hypothetical flare blast wave.
These circumstances along with the framework of the ``flare vs.
CME'' alternative probably account for the conclusions made in some
case studies in favor of flare-ignited shock waves. On the other
hand, this alternative and observational difficulties might incline
different studies toward the initial bow-shock excitation by the
outer surface of a super-Alfv{\'e}nic CME.

Being constrained by these difficulties, researchers are forced to
invoke indirect arguments, which do not always ensure the
unambiguous identification of a scenario. These are, for example,
the presence of a fast CME that cannot guarantee the bow-shock
regime of an associated wave. It is also not certified by the
position of the Type~II source ahead of a CME, because the Type~II
emission can originate from the streamer above the eruption region
disturbed by the quasi-parallel blast-wave-like shock. Next, a
delayed appearance of a Type~II burst that does not necessarily mark
the onset of the shock formation. On the other hand, the absence of
a CME is not evidence of the flare-related shock origin, as
mentioned.

\subsection{Overview of Actual Shock-Wave Histories}

To avoid deceptive indications, it is reasonable to follow the
appearance and evolution of shock waves and to measure their
propagation from a combined analysis of their various manifestations
in different spectral ranges. This way is time-consuming, but
provides a highest confidence in adequacy of the outcome. Using this
approach, we made a detailed analysis of the shock-wave histories
for several events in a manner similar to Section~\ref{S-march16},
mainly from the extreme-ultraviolet and white-light coronagraph
images, dynamic radio spectra, and others (e.g. H$\alpha$ images),
if available. The results of these case studies are summarized in
Table~\ref{T-summary}, whose column 15 specifies the article, where
they were published.

Table~\ref{T-summary} contains 13 events listed chronologically.
The kinematics of eruptive filaments or similar structures was
measured in 8 events, when it was possible. Two shock waves
following each other and merging eventually into a single stronger
shock were revealed in four events. Column 1 lists the number of
an event with a label ``a'' or ``b'' specifying one of the two
shocks, if present. Columns 2--5 list the date (in the format of
the Solar Object Identifier), peak time, duration, and importance
of a flare according to the GOES reports, and column 6 gives its
reported position. Columns 7--9 present the estimated wave onset
time, the peak time of an HXR or microwave burst, and the onset
time of a Type~II burst. Columns 10--12 present the CME parameters
taken from the online CME catalog
(\url{https://cdaw.gsfc.nasa.gov/CME_list/}:
\citealp{Yashiro2004}): the onset time at the limb estimated from
a linear fit and second-order fit, and an average speed. Column 13
shows the outcome of the shock-wave history: either a bow shock,
or decay. Column 14 lists the peak flux of near-Earth protons $>
10$\,MeV produced by the event (GOES).

\begin{table*}
 \footnotesize
\centering \caption{Summary of shock waves studied}
\label{T-summary}
\begin{tabular}{lllrcclccccrcrc}

\hline \noalign{\vskip 1mm}

\multicolumn{1}{c}{No.} & \multicolumn{1}{c}{Date} &
\multicolumn{3}{c}{GOES} & \multicolumn{1}{c}{Position} &
\multicolumn{1}{c}{Wave} & \multicolumn{1}{c}{$T_\mathrm{peak}$} &
\multicolumn{1}{c}{Type II} & \multicolumn{3}{c}{CME} &
\multicolumn{1}{c}{Shock} & \multicolumn{1}{c}{$J_{10}$} & \multicolumn{1}{c}{Refs}\\

\cline{3-5} \cline{10-12} 
\multicolumn{2}{c}{} & \multicolumn{1}{c}{Peak} &
\multicolumn{1}{c}{Dur.} & Size & \multicolumn{1}{c}{} &
\multicolumn{1}{c}{onset} & \multicolumn{1}{c}{HXR or} &
\multicolumn{1}{c}{onset} & \multicolumn{2}{c}{Onset at
$1\,\mathrm{R}_\odot$} & \multicolumn{1}{c}{Speed} & \multicolumn{1}{c}{outcome} & \multicolumn{1}{c}{[pfu]} \\

\multicolumn{1}{c}{} & \multicolumn{1}{c}{} &
\multicolumn{1}{c}{time} & \multicolumn{1}{c}{min} &
\multicolumn{1}{c}{} & \multicolumn{1}{c}{} &
\multicolumn{1}{c}{time} & \multicolumn{1}{c}{m/w} &
\multicolumn{1}{c}{time} & 1-order & 2-order  &
\multicolumn{1}{c}{km\,s$^{-1}$} &
\multicolumn{1}{c}{} &  \\

\hline 

\multicolumn{1}{c}{1} & \multicolumn{1}{c}{2} &
\multicolumn{1}{c}{3} & \multicolumn{1}{c}{4} &
\multicolumn{1}{c}{5} & \multicolumn{1}{c}{6} &
\multicolumn{1}{c}{7} & \multicolumn{1}{c}{8} &
\multicolumn{1}{c}{9} & \multicolumn{1}{c}{10} &
\multicolumn{1}{c}{11} & \multicolumn{1}{c}{12} & \multicolumn{1}{c}{13} &
\multicolumn{1}{c}{14} & \multicolumn{1}{c}{15}\\

\hline \noalign{\vskip 1mm}

1  & 1997-09-24 & 02:48:00      & 9 & M5.9            & S31E19   &
02:46:50        & 02:46:50 & 02:48:40           & 02:33 &
00:55$^{1}$    & 532                 & Decay & -- & 1               \\
2a & 2001-12-26 & 05:40:00      & 135                    & M7.1 &
N08W54   & 05:04:00        & 05:04:40                        &
05:08:00           & 05:06 &
05:10      & 1446                & Bow & 700          & 2    \\
2b &            & & & & & 05:09:00 & 05:09:00
& 05:12:00 & \multicolumn{3}{c}{--------- Same ---------} & \multicolumn{3}{c}{------ Same ------} \\
3  & 2002-06-01 & 03:57:00      & 11                     & M1.5 &
S19E29   & 03:53:40        & 03:53:40                        &
03:55:30           & \multicolumn{3}{c}{No coronagraph data}  &
Decay?
& --         & 1       \\
4 & 2003-11-18 & 07:52:00      & 43                     & M3.2 &
N00E18   & 07:41:00        & 07:42:00                        &
07:47:00           & \multicolumn{3}{c}{Confined eruption}
                   & Decay &   --    &   3         \\
5 & 2003-11-18     & 08:31:00      & 47 & M3.9 & N00E18 & 08:14:12
& 08:16:00                        & 08:15:00 &
08:13 & 08:13      & 1660                & Bow & 0.7  & 3            \\
6  & 2004-07-13 & 00:17:00      & 14                     & M6.7 &
N13W46   & 00:14:50        & 00:15:00                        &
00:16:00           & 00:02 &
00:04      & 607                 & Decay & 1    & 1,4            \\
7a & 2006-12-13 & 02:40:00      & 43                     & X3.4 &
S06W23   & 02:23:20        & 02:25:30                        &
02:26:00           & 02:25 & 02:29      & 1774                & Bow & 695   & 5           \\
7b &            & &                        &                 & &
02:27:20        & 02:29:00 & 02:28:00$^2$&
\multicolumn{3}{c}{--------- Same ---------} & \multicolumn{3}{c}{------ Same ------}                  \\
8a & 2007-05-19 & 13:02:00      & 31                     & B9.5 &
N07W06   & 12:50:00        & 12:51:15                        &
12:52:00           & 12:56 & 13:00      & 958                 & Decay & --    & 1            \\
8b &            & &                        &                 & &
12:56:00        & 12:57:00                        & 13:01:00
&    \multicolumn{3}{c}{--------- Same ---------} & \multicolumn{3}{c}{------ Same ------}              \\
9  & 2010-01-17 & 03:56:00      & ? & X1$^{3}$       & S25E128 &
03:47:48        & No data                         & 03:51:00 &
03:13 & 03:45      &
350                 & Decay     &         +      & 6   \\
10  & 2010-06-13 & 05:39:00      & 14                     & M1.0 &
S21W82   & 05:35:10        & 05:36:00                        &
05:38:00           & 05:14 &
04:56$^{1}$    & 320                 & Decay & --     & 7           \\
11 & 2011-02-24 & 07:35:00      & 19                     & M3.5 &
N19E84   & 07:29:00        & 07:30:30                        &
07:34:30$^{4}$         & 07:16 &
07:23      & 1186                & Bow & --       &     8     \\
12 & 2011-05-11 & 02:43:00      & 60                     & B8.1 &
N25W54   & 02:22:10        & 02:28:30$^{5}$ & 02:27:00 &
02:26 & 02:24      & 745                 & Decay & 0.5  & 8            \\
13 & 2016-03-16 & 06:46:00      & 23                     & C2.2 &
N14W83   & 06:36:30        & 06:37:30                        &
06:45:00           & 06:04 & 06:21      & 592 & Decay & 1 & 9 \\
\hline
\end{tabular}

\flushleft

$^{1}$ Acceleration is uncertain due to either poor height
measurement or a small number of height-time measurements (remark
from the CME catalog).

$^{2}$ Not clear.

$^{3}$ Average of the estimates from STEREO-B/EUVI 195\,\AA\
images of M6.4 by \cite{Nitta2013_farside} and X1.6 by
\cite{Chertok2015}.

$^{4}$ Reported 07:37:00 when the Type II structures became clear
after overlap with a strong Type III group.

$^{5}$ For the derivative of the GOES flux at 1--8\,\AA.

References: 1.~\cite{Grechnev2011_I}, 2.~\cite{Grechnev2017_III},
3.~\cite{Grechnev2014_II}, 4.~\cite{Grechnev2008},
5.~\cite{Grechnev2013_6dec}, 6.~\cite{Grechnev2011_III},
7.~\cite{Grechnev2016}, 8.~\cite{Grechnev2015}, 9.~Present
article.

\end{table*}

The events listed in Table~\ref{T-summary} had greatly differing
properties. The flares ranged in size from B8.1 to X3.4 and in
duration from 9 to 135 minutes. The average CME speed ranged from
320\,km\,s$^{-1}$ to 1774\,km\,s$^{-1}$. Noteworthy was event 4,
in which a confined eruption without any CME produced a shock
wave, which excited clear large-amplitude oscillations of a remote
filament observed in the H$\alpha$ line center and both wings
(``winking filament''). The flares in the 13 events had differing
morphologies, including two-ribbon flares and flares with circular
ribbons. Nevertheless, the shock-wave excitation scenario was the
same in all of these events. The wave onset times were close to
the peak times of the HXR or microwave bursts or led them by up to
2 minutes (when they were observed), i.e. occurred not later than
the flare impulsive phase.

Despite the differences between the events listed in
Table~\ref{T-summary}, shock waves in all of them were initially
excited in the same impulsive-piston scenario by sharply erupting
filaments or similar structures, as described in
Section~\ref{S-EUV_wave}. This fact allows combining the results
obtained in studies of different events to reveal common properties
of these shock waves. The possibility of their flare-related origin
was examined in each case study and excluded for the reasons listed
in Section~\ref{S-scenarios}.

Neither was a shock initially excited in any of the events by a
super-Alfv{\'e}nic CME. This result is also expected, because the
impulsive-piston shock excitation by a relatively small erupting
structure is highly efficient in a medium with a steep falloff of
the fast-mode speed away from the eruption region. Hence, the
shock appears much earlier than is possible in the bow-shock
scenario; the shock waves initially resemble blast waves. While
they eventually changed to the bow-shock regime in 4 events in
Table~\ref{T-summary}, this did not affect their early
development. The successive appearance in events 2, 7, and 8 of
two shock waves within 6 minutes supports this conclusion, because
a single super-Alfv{\'e}nic CME cannot drive more than one shock.

The initial wave excitation and the CME development turn out to be
closely related. Most likely, when an eruption starts, neither a
CME nor its flux rope exists in the final form. For example, wave
traces in event 10 were revealed inside the developing CME; then
the wave passed through its structures and propagated outward like
a decelerating blast wave \citep{Grechnev2016}. There is no reason
for a concern about the role in the shock-wave excitation of a
presumable lateral overexpansion of the CME bubble, which does not
yet exist at that time. There was nothing to expand laterally in
event 13 (Section~\ref{S-march16}); nevertheless, the shock wave
appeared.

The CME speeds listed in column 12 of Table~\ref{T-summary} are
related to the plane of the sky, while the CME orientations could be
strongly off-plane. The speeds might therefore be underestimated
considerably for slow CMEs and moderately for fast CMEs, whose
measurements are probably related to nearly spherical wave fronts.
With these circumstances, the transition to a CME-driven shock
occurs for those CMEs, whose average speed exceeds
1000\,km\,s$^{-1}$. Indeed, to ensure the super-Alfv{\'e}nic regime,
the CME speed should exceed the sum of the Alfv{\'e}n speed and the
solar wind speed. Using the models of the Alfv{\'e}n speed
\citep{Mann2003} and solar wind speed \citep{Sheeley1997},
\cite{Grechnev2017_III} estimated this sum to decrease from
900\,km\,s$^{-1}$ at $5\mathrm{R}_\odot$ to 650\,km\,s$^{-1}$ at
$25\mathrm{R}_\odot$ (with an established solar wind speed of
400\,km\,s$^{-1}$). Nevertheless, with a CME speed as high as
1446\,km\,s$^{-1}$ in event 2, the bow-shock regime became possible
at distances exceeding $15\,\mathrm{R}_\odot$, while the wave front
was still nearly spherical \citep{Grechnev2017_III}.

The transition of a blast-wave-like shock to a CME-driven bow
shock corresponds to the change from the regime of the plasma
extrusion by the CME body to the regime of the plasma flow around
its outer surface, when the aerodynamic drag becomes significant.
This change occurring at considerable distances from the Sun
determines the shape of a CME-driven shock. It forms from a nearly
spherical blast-wave-like shock, while its driver expands in three
dimensions \citep{VrsnakCliver2008, Grechnev2011_I}. This makes
the bow-shock shape with a Mach cone unlikely and raises a
question about its actual shape. An additional consequence of
Table~\ref{T-summary} is the early shock-wave appearance in events
2 and 7 responsible for major energetic particle events and GLE63
and GLE70. This circumstance should be considered in studies of
solar energetic particles.

All of the listed events were associated with decelerating shock
waves. The drag should also decelerate fast CMEs, when it becomes
important. These circumstances imply that the onset time of a
corresponding transient estimated from the second-order fit should
generally be somewhat later than that estimated from the linear
fit. This pattern mostly holds for the events listed in
Table~\ref{T-summary}, except for those whose observations were of
an insufficient quality (events 1, 10 and 12; they were equal for
event 5). A positive acceleration estimated in the CME catalog for
fast CMEs is probably a result of observational difficulties.

Besides the implications mentioned, there are several other
significant consequences of the shock-wave histories discussed.
All of them emphasize the importance of systematic studies of
coronal shock waves. Statistical studies of EUV waves have
recently been made by \cite{Nitta2013_waves}, \cite{Muhr2014}, and
\cite{Long2017}. Some of their conclusions do not agree with each
other, probably because of the observational difficulties shown in
Sections \ref{S-EUV_wave} and \ref{S-scenarios}. Some others do
not seem to be obvious. Our results can shed light on these
challenges.

For example, all of these studies stated a poor correspondence
between EUV waves and Type~II bursts. This seems to be
challenging, if the Type~II emission originates ahead of a CME,
while \cite{Muhr2014} consider them as the EUV waves' driving
agent. The situation is different, if Type~IIs originate in
streamers located away from the eruption region. Such a streamer
may exist or may not. If the antiparallel magnetic fields in a
streamer are separated by plasma outflow caused, e.g., by a
preceding CME, then the streamer cannot generate Type~II emission.
On the other hand, the visibility of an EUV wave is determined by
the ambient fast-mode speed and can be poor, e.g., in coronal
holes \citep{Grechnev2011_III, Long2017}. The plasma density
depletion caused by a preceding CME also disfavors the detection
of an EUV wave. These circumstances might be implicated in the
extreme cases of mismatch between EUV waves and Type~II bursts
shown by \cite{Nitta2014}.

The pattern found by \cite{Muhr2014} and \cite{Long2017}, with
faster EUV waves exhibiting a stronger deceleration, suggests that
the highest-speed initial stage of the EUV wave propagation is often
not fully measured, as the velocity--time plot in
Figure~\ref{F-aia_wave_kinem}b explains. Some causes of a poor EUV
wave visibility are mentioned in the preceding paragraph.

The absence of any relationship between the EUV wave properties and
the size of the associated flare stated by \cite{Nitta2013_waves}
and \cite{Long2017} is consistent with our results. Instead, the
shock-wave excitation mechanism we are talking about is expected to
depend on the acceleration of an eruptive structure that is not easy
to measure.

\subsection{The Role of the Flare Duration in Soft X-rays}

There is a traditional view relating impulsive flares to narrow or
no CMEs and long-decay flares (LDEs) to large CMEs
\citep{Kahler1989}. While the authors of this statement talked
primarily about major flares ($\geq$\,M1 GOES importance), this
pattern obviously holds for minor events presented in
Section~\ref{S-neg_bursts}. However, a wide CME on 16 March 2016
discussed in Section~\ref{S-march16} developed also in association
with an impulsive flare. Some CMEs in Table~\ref{T-summary} were
also related to impulsive flares. \cite{NittaHudson2001} presented
a series of large CMEs, which occurred in association with major
impulsive flares in the same active region within 60 hours.
Conversely, infrequent major LDEs without any eruptions are known
(e.g. \citealp{Thalmann2015}). Thus, the pattern found by
\cite{Kahler1989} seems to represent a tendency, but does not
ensure a one-to-one correspondence.

The long decay time in LDEs might be determined by long-lasting
reconnection processes occurring typically in the post-eruption
phase \citep{Grechnev2006} or at a late stage of rare confined
flares. The conditions favoring such processes still need
understanding.

On the other hand, the SXR GOES fluxes might possibly be invoked
to find the indications of a probable EUV wave occurrence.
According to the Neupert effect, the rise time of the SXR flux
should correspond to the acceleration duration of an eruption.
Being possibly somehow combined with another parameter of an
event, this rise time might characterize its impulsiveness to
indicate the magnitude of the acceleration and thus to provide an
indication of a probable EUV wave.

\section{Summary and Conclusion}
\label{S-summary}

The T-shaped SRH antenna array with redundant baselines has
allowed implementing algorithms to construct correlation plots of
the solar radio emission and those to synthesize the images of the
Sun without involvement of calibration radio sources. A high
sensitivity of the interferometer of about 0.01\,sfu in
combination with a high dynamic range makes it possible to observe
in microwaves without attenuators a wide range of solar activity,
from sources of powerful flare bursts down to its faint
manifestations associated with microeruptions. The latter occur
more frequently, being less studied.

The first observations with SRH have shown its promising
opportunities to detect solar eruptions of different energy and
spatial size. We have demonstrated three ways to detect the
eruptions: i)~direct observations of erupted material,
ii)~observations of microwave bursts as a probable pointing at
eruptive events, and iii)~detection of faint eruptive events that
manifest as depressions in the total-intensity correlation plots,
being accompanied by distinct changes in the circular-polarization
plots. Such events can be too weak and small to be detected from
any other observations. We have learned from the SRH observations
that microwave depressions at 4--8\,GHz of this kind are typically
polarized. They can be caused by eruptions from the same region
repeating in a few hours, and this can occur not once. Such
phenomena raise a question what favors energy release in small
portions, preventing its accumulation. An answer might shed
additional light on preparation conditions and their
manifestations for big eruptions which pose a largest space
weather hazard. Understanding of the mechanisms responsible for
the eruptions of different size, their implication to space
weather disturbances as well as development of criteria for their
detection is among important future tasks for the multi-frequency
SRH.

To carry out detailed studies of solar eruptions, it is reasonable
to combine the SRH observations with multi-instrument data from
different spectral ranges. This is a typical approach in such
studies. Besides the listed opportunities to detect various
eruptions, a significant advantage of the SRH observations is
promised by their dense frequency sampling: in December 2017, the
SRH has started observing first at 15 frequencies, and then at 32
frequencies within the 4--8\,GHz range. In February 2018, the time
to process each frequency bin has been reduced and reached a
planned value of 0.28\,s. The time to collect the visibilities at
32 frequencies became about 9\,s.

From the multi-instrument analysis of an eruptive event observed
by the SRH on 16 March 2016, we have followed the development of a
CME and associated shock wave and compared them with expectations
from well-known models. This event has demonstrated a direct
shock-wave excitation by an erupting prominence without any
indications of a cavity or rim bounding it that contradicts their
crucial role presumed in some studies. Another highlight of this
event is that the shock wave, which was probably responsible for a
near-Earth proton enhancement, was not CME-driven and appeared
during the flare impulsive phase, when the CME was still in the
development stage. Thus, a widely accepted view on the origin of
solar energetic particles should be refined. The scenario
discussed appears to be typical of various solar eruptions of
different importance. We hope our results would be helpful in
further studies of solar eruptions, CMEs, and coronal shock waves.

\section*{Acknowledgements}

We thank N.V.~Nitta and H.~Nakajima for fruitful discussions and
I.V.~Kuzmenko for the RT-2 data. We appreciate our colleagues from
the Radio Astrophysical Department and the Radio Astrophysical
Observatory in Badary. We are indebted to anonymous reviewers for
their valuable remarks. We thank the NASA/SDO and the AIA and HMI
science teams; the instrument teams of GOES, RHESSI, the SWAP
telescope on the ESA's PROBA2 spacecraft, the NASA's Fermi
Gamma-Ray Space Telescope, Culgoora and Learmonth spectrographs of
the Australian Space Weather Services, and LASCO on SOHO. SOHO is
a project of international cooperation between ESA and NASA. We
thank the team maintaining the CME Catalog at the CDAW Data Center
by NASA and the Catholic University of America in cooperation with
the Naval Research Laboratory.

The work was performed with budgetary funding of Basic Research
program II.16. The results were obtained using the Unique Research
Facility Siberian Solar Radio Telescope
\url{http://ckp-rf.ru/usu/73606/}.

  \bibliographystyle{elsarticle-harv}

\bibliography{jastp_lit}

\end{document}